\documentclass{article}
\usepackage{graphicx} % Required for inserting images
\usepackage{geometry}
\usepackage{afterpage}
\usepackage{bm}
\usepackage{graphicx}
\usepackage{float}
\usepackage{amsmath}
\usepackage{pgf, tikz}
\usetikzlibrary{decorations.markings}
\usetikzlibrary{patterns}
\usetikzlibrary{math}
\usepackage{amsthm,amsfonts,amssymb,stmaryrd}
\usepackage{hyperref}
\usepackage{tcolorbox}
\usepackage{url}
 \usepackage{algorithm}
\usepackage{algorithmic}
\usepackage{adjustbox}
\usepackage{array}
\usepackage{booktabs}
\usepackage{multirow}
\usepackage{nicematrix}
\usepackage{diagbox}  
\usepackage{lmodern}
\newtheorem{re}{Remark}[section]
\newtheorem{hyp}{Assumption}

\newtheorem{pb}{Problem}[section]

\newtheorem{theo}{Theorem}[section]
\newtheorem{lemma}{Lemma}[section]
\newtheorem{coro}{Corollary}[section]

\newtheorem{prop}{Proposition}[section]
\newtheorem{defi}{Definition}[section]

\newcommand{\hquad}{\hspace{1.8mm}}
\newcommand{\ftilde}{\widetilde{\mathbf{f}}}
\newcommand{\f}{\mathbf{f}}

\newcommand{\G}{\mathcal{G}}
\newcommand{\GQ}{\mathcal{G}_Q}
\newcommand{\Go}{\mathcal{G}_o}

\newcommand{\Ho}{\mathcal{H}_o}

\newcommand{\Goohat}{\widehat{\mathcal{G}_{oo}}}

\newcommand{\Goo}{\mathcal{G}_{oo}}
\newcommand{\jac}{\textrm{jac}}

\newcommand{\rank}{\textrm{rank}}
\newcommand{\crit}{\textrm{crit}}
\newcommand{\sing}{\textrm{sing}}
\renewcommand{\gcd}{\textrm{gcd}}
\newcommand{\signat}{\textrm{signature}}
\newcommand{\n}{\mathfrak{n}}
\newcommand{\sign}{\textrm{sign}}

\newcommand{\ov}[1]{\overline{#1}}
\newcommand{\wt}[1]{\widetilde{#1}}
\newcommand{\wh}[1]{\widehat{#1}}
\newcommand{\cc}[1]{\pi_0(#1)}
\newcommand{\ccpetit}[1]{\pi_0(#1)}
\newcommand{\CC}{\mathcal{C}\hspace*{-3pt}\mathcal{C}}
\newcommand{\W}{\mathcal{W}}
\newcommand{\set}[1]{\mathcal{Z}(#1)}
\newcommand{\VR}[1]{\mathbf{V}_{\mathbb{R}}(#1)}

\usepackage[symbol]{footmisc}
\title{\textbf{A certified classification of first-order controlled coaxial telescopes.}}

\author{Audric Drogoul\\
\small Thales Alenia Space\\
 \small 5 All\'ee des Gabians, 06150 Cannes, FRANCE.\\
  \small\texttt{audric.drogoul@thalesaleniaspace.com}
}
\date{\small \today}
\usepackage{appendix}
\usepackage{fancyhdr}
\usepackage{fbox}
\definecolor{tas_lim_green}{RGB}{0, 171, 70}

\newcommand{\TASLim}{
\begin{minipage}[t]{0.9\textwidth}
    \raggedright % Aligner à gauche le texte
    \tiny
    \vspace{0pt}
    \textbf{Property.}
    This document is not to be reproduced, modified, adapted, published translated in any material form in whole    or in part nor disclosed
     to any third party without the prior written permission of Thales Alenia Space.\\
    © 2024 Thales Alenia Space all rights reserved.  \\
    
  \end{minipage}

}

\fancypagestyle{plain}{

  \fancyhead{}

  \fancyhead[L]{\thepage}

  \fancyhead[R]{\nouppercase{\leftmark}}

  \fancyfoot{}

   \fancyfoot[C]{\TASLim}
}
\usepackage{cleveref}
\crefname{re}{Remark}{Remark}
\crefname{hyp}{Assumption}{Assumption}
\crefname{pb}{Problem}{Problem}
\crefname{proc}{Process}{Process}
\crefname{ex}{Example}{Example}
\crefname{theo}{Theorem}{Theorem}
\crefname{lemma}{Lemma}{Lemma}
\crefname{coro}{Corollary}{Corollary}

\crefname{prop}{Proposition}{Proposition}
\crefname{defi}{Definition}{Definition}
\begin{document}
\pagestyle{plain}
\newgeometry{top=0.2cm} % Ajustez la marge supérieure
{

\makeatletter
\addtocounter{footnote}{1} % to get dagger instead of star
\renewcommand\thefootnote{\@fnsymbol\c@footnote}%
\makeatother

\maketitle
}

% REQUIRED
\begin{abstract}
    This paper is devoted to an intrinsic geometrical classification of three-mirror  telescopes. 
    The problem is formulated as the study of the connected components of a semi-algebraic set.
      Under first order approximation, we give the general expression of the transfer matrix of a reflexive optical system.
      Thanks to this representation, we express the semi-algebraic set for focal telescopes and afocal telescopes as the set of non-degenerate  real solutions of first order optical conditions.
      % for focal telescopes, we  express  focal, null Petzval's curvature and telecentricity conditions as polynomials equations depending on the inter-mirror distances and mirror magnifications.
      % Eventually, the set of admissible focal telescopes is written as real solutions of aforementioned polynomial equations under non degenerating conditions that are non-null curvatures and non-null magnifications.
      % The set of admissibile  afocal telescopes is written analogously.
      Then, in order to study the topology of these sets,  we address the problem of counting and describe their connected components.
      In a same time, we introduce a topological invariant which encodes the topological features of the solutions.
      For systems composed of three mirrors, we give the semi-algebraic description of the connected components of the  set and show that the  topological invariant is exact.

      % In a same time, we introduce an exact topological invariant which encodes the topological features of the solutions.
      % To achieve this, we consider the canonical projection on a well-chosen parameter space and we split the semi-algebraic set w.r.t  the locus of the critical points of the projection restricted to this set.
      %  Then, we  show that each part  projects homeomorphically for $N=3$ and we obtain the connected components of the initial set by merging those of each part through the set of critical points of the introduced projection.
      %  Besides, in that case, we give the semi-algebraic description of the connected components of the initial  set and introduce a topological invariant and a nomenclature which encodes the invariant topological/optical features of optical configurations lying in the same  connected component.
  
  \end{abstract}

  \section{Introduction}
  %%context
Optical designing is a scientific and engineering discipline performed by experimented opticians,
where the goal is often to construct an  optical
system that optimizes   optical, geometrical and manufacturability criteria. 
During the designing process, opticians manly focus on geometrical and optical performances and check the manufacturability and stability to misalignment at posteriori.
Generally the design exploration is split in several steps which gradually converge to the target solutions  as discussed in \cite{Rolland18,Papa:21}.
A first step consists in neglecting the obscuration and  considering on-axis  conic-based solutions which enjoy to a rotationally symmetry  cancelling the aberrations of even orders. 
First orders equations fix the curvatures while the third-order rotationally invariant Seidel aberrations can be corrected by  conics \cite{Korsch:91}.
Then the system can be unobscured by tilting the surfaces and using a combination of field-bias and offset aperture. 
This latter step generally introduces  rotationaly variant aberrations which can be corrected by additional degree of freedom on the shape of the optical surfaces which takes the name of \textit{freeforms}. 
As explained in \cite{Rolland18}, the introduction of freeforms is not always sufficient to correct the optical aberrations and a large increase in freeform departure for each surface
can be associated to a little performance gain. Let us note that  the more the freeform departure is high the more the fabrication time is high and so solutions with few freeform sag are preferred. 
This is why the choice of a good starting point before introducing freeforms is important and in particular the choice of distances and curvatures, which determines conics by  linear relations  \cite{Korsch:91}, can be crucial for the sequel of the process.
% In particular, non symmetric shape corrections, such as Zernike polynomials, are used to model a Freeform optical surface.
\newpage \restoregeometry\indent This paper addresses the study of \textit{admissible} on-axis optical configurations   which are real solutions of a set of first order equations  determining for example the curvatures of the system given
 \noindent inter-mirror distances. 
In our case, an optical configuration is \textit{admissible} if it contains no flat surfaces and if no intermediate magnification is zero (which would correspond to a zero surface size).
Generally, the optical designer loops on a thousand of \textit{admissible} on-axis configurations among which he hopes to find the one it will converge, after applying the above steps, to an admissible unobscured and aberration less feasible solution.
Besides, after this first step,  curvatures and distances satisfying  focal or magnification constraints are not changed anymore so that each configuration  verifies a set of first order equations that we want to preserve by correcting the optical aberrations during the following process.
However, among this huge amount of solutions, a lot are optically similar and no guarantee of completeness is provided.
Hence, understanding the geometry of the solution set associated to classical first order equations is a very important question.

In  litterature, classification appears as an open question linked to the understanding of optical design methods. 
For two-mirror systems, \cite{2020Trumper} proposes a methodology for classifying obscuration-free solutions unfolded in the plane.
Two classes are heuristically identified, omitting the $\texttt{VAVA}$ class presented in \cite{offaxis24}.
In \cite{Rolland18}, three mirror co-axial telescopes  are classified by considering only the signs of the mirrors' curvatures using names like $\texttt{PNP}$ to states that the first mirror is convex the second is concave and the last one is convex. 
    As we will see, to classify such telescopes described as  an affine variety satisfying  a set of first-order conditions includig the focal one, the signs of the mirrors' curvatures   do not constitute an exact invariant. 
    As last example, \cite{Papa:21}  classifies four-mirrors based configurations by considering the presence of internal intermediary images and pupils.

    As we can see, the divergent ways of classifying optical configurations testify to the need to reformulate the question mathematically.
     The mathematical question that we propose to answer is  counting and describing  the connected components of this solution set and introducing a topological invariant  (see Definition~\ref{de:topologicalInvariant}) and a nomenclature (Definition~\ref{def:nomenclature}) that intelligibly encodes this topological invariant.  
Hence thanks to this meaningful nomenclature, opticians can draw the main features of the optical configuration by just knowing its name.
Let us emphasize that the set of connected components of a set are the equivalence classes in the sense of the homotopy relation of that set. 
Hence, optical configurations lying in the same  connected component are equivalent by a continuous deformation.
This fact is crucial for the continuation of the optical design process, where optical configurations are continuously deformed by a gradient flow of a certain cost function. 
Hence answering the aforementioned  question enables  to (i) understand the optical/topological invariance of the on-axis optical configurations inside the classes encoded in an intelligibly nomenclature (ii)  mathematically certify that the all classes are represented.
Let us note that a similar approach to classify off-axis obscuration free solutions is  developed  in \cite{offaxis24} where authors introduce an off-axis mathematically certified nomenclature which can be used with the on-axis present one to get a complete on/off-axis nomenclature. 
\\
\indent To achieve this, we introduce the set of first order equations thanks to a transfer matrix formalism \cite{PerezBook} and we explicit them for optical configurations composed  of $N$ mirrors in function of distances inter-mirrors and magnifications. 
Expressing these equations as polynomials ones, the set of \textit{admissible} solutions writes as a semi-algebraic set, what enables to use powerful mathematics tools of the domain of real algebraic geometry and computer algebra as done in others engineering disciplines such as Robotics \cite{CapcoEldin} or Biology \cite{chauvin}.
We show that the set of equations satisfied for $N=3$ mirrors can be written as a triangular system with a parametric trinomial  as a pivot equation plus an equation fixing the product of the unknowns which leads to  finite fibers every where on the parameter admissible space.
Inspired by the real root classification algorithm \cite{LE202225} which gives a way to construct explicitly homeomorphisms between  a dense partition of an algebraic set and a dense subset of its canonical projection $\pi:(\mathbf{y},\mathbf{x})\in \mathbb{R}^{t+n}\to \mathbf{y}\in\mathbb{R}^t$ with $t$ the dimension of the real algebraic set.
In particular, we decompose the set into two parts separated by the set of critical points of $\pi$, we show that each is homeomorphic to its projection what corresponds to the main resut stated in \cref{theo:xi1epsContinuousThroughA1epsB1} and \cref{theo:CCdeGepsetCCdeE}.
Next, by linking the connected components of each part to those of the initial set, we deduce the connected components of the initial set by merging the obtained components  through the set of critical points of $\pi$.
%Then we establish a link between the connected components of each part and the connected components of the initial set by merging the former over the set of critical points of $\pi$.
The steps of the connected components computation are summarized in \cref{alg:computeGeps}. \newline
% Besides, exploiting the fact that the product of magnifications cannot cancel we succeed to construct two homeomorphisms defined between a partition of the semi-algebraic and its associated projection which enables to characterize the connected components of 
% the initial semi-algebraic from its projected subspaces.
\indent The paper is organized as follows. In \cref{sec:polynomialsSystems} we introduce the considered first order equations for focal and afocal telescopes, in \cref{sec:pb_classif} we introduce the classification problem and in particular we start by \cref{sec:preliminaries} by introducing some preliminaries of real algebraic
 geometry and we use it to study the connected component of a generic triangular system with parametric trinomial as
%  \newpage\phantom{ }\vspace{-3cm} 
%  \noindent 
 pivot equation  whose the product of the unknown cannot cancel in \cref{sec:triangular_system}.  We summarize the step of this computation  in \cref{alg:computeGeps}.
 Let us note that the first step of this algorithm consists in performing a real root classification which  is done in \cite{LE202225} for a general polynomial system.
In \cref{sec:optical_application} we apply the previous result to give a name (see \cref{def:nomenclature}), a semi-algebraic representation and a sample point for each connected component of the admissible solutions set and a graphical representation are given.

  \section{Polynomial systems\label{sec:polynomialsSystems}}
  This section is a short review of first order optics from which we derive the polynomial system we propose to study in the paper. We start by intoducing a parametrization of the curvatures depending on the distances between mirrors and focal plane and  the lateral magnification of the mirrors.
  Then we specialise the polynomial system for focal (resp. afocal) telescopes where we give the expression of the polynomials associated to the focal (resp. magnification) condition, null Petzval curvature condition and telecentricity (resp. exit pupil position w.r.t entry pupil position) constraint.
  
  \subsection{Problem statement}
  \label{sec:pb_statement}
\begin{figure}[H]
    \begin{center}
        \includegraphics[width=0.7\textwidth]{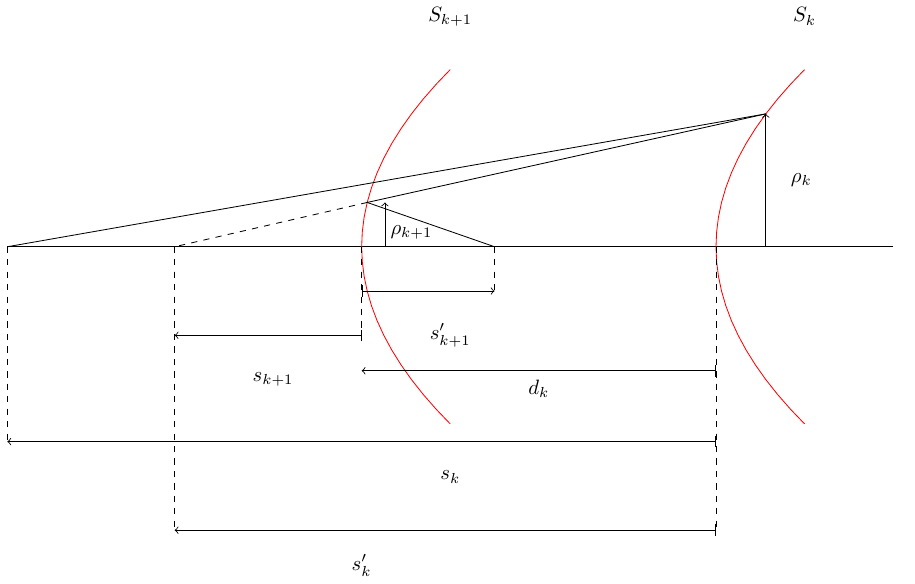}
    \caption{\label{fig:notationKorsh}Geometrical illustration of \eqref{eq:courbures}. 
    First order formula: $\frac{1}{s_k}+\frac{1}{s_k'}=2c_k$, Change of coordinate system: $d_k=s_k'-s_{k+1}$, Magnification definition: $\Omega_k = \frac{\rho_{k+1}}{\rho_k}=\frac{s_{k+1}}{s_k'}$}
    \end{center}
\end{figure}

  Let  $N$ be the number of mirrors, $S_k$ be the $k$-th mirror, $c_k$ be its curvature for $1\leq k \leq N$, and   $d_k$ the signed distances between $S_k$ and $S_{k+1}$ relatively to increasing $z$ with the convention that $S_{N+1}$ is the focal plane (possibly at infinity). 
By denoting $v_1$ the inverse of the distance between the observed object and the first mirror $S_1$, we can deduce  the position of its image after reflecting the first mirror by a first order formula: $v_1+\frac{1}{s_1'}=2c_1$ with $s_1'$ is the first intermediate image position w.r.t the center of $S_1$ and relatively to increasing $z$. 
Re-expressing $s_1'$ in the coordinate system of $S_2$ enables to define $s_2=s_1'-d_1$ and re-imaging $s_2$ by the second mirror $S_2$ gives $s_2'$ with $\frac{1}{s_2}+\frac{1}{s_2'}=2c_2$. 
Repeating this procedure  $N$ times and using the magnification definition of the $k$-th mirror $\Omega_k = \frac{s_{k+1}}{s_k'}$ (see \cref{fig:notationKorsh}) gives the following curvatures expression: 
 
\begin{equation}
\begin{array}{ccc}
    c_1 = \frac{(1-\Omega_1)}{2d_1}+\frac{v_1}{2},\hquad &
    c_k = \frac{(1-\Omega_{k-1})}{2\Omega_{k-1}d_{k-1}}+\frac{(1-\Omega_k)}{2d_k}\text{ for }k\in \llbracket 2,N-1\rrbracket,\hquad 
    c_{N} =\frac{(1-\Omega_{N-1})}{2\Omega_{N-1}d_{N-1}}+\frac{v_N'}{2}\\
    \end{array}
    \label{eq:courbures}
\end{equation}
where  $v_1=0$ if the object is at infinity and $v_N'=\frac{1}{d_N}$.\\
% Ajustez la marge supérieure
\paragraph{Notations}
 Let $X=(X_1,\dots,X_l)$ be a real finite sequence of length $l$ and $I\subset \llbracket 1,l\rrbracket$, we denote $X_s=\prod_1^lX_i$ and $\widehat{X_{I}}=\prod_{i\in\llbracket 1,\rrbracket\backslash I}X_i$.
% \begin{align*}
%     \Omega_{s}=\Omega_{s_{N-1}} =\prod_1^{N-1}\Omega_k,\hspace{3mm}
%     &d_{s} =d_{s_N}=\prod_1^N d_k,\hspace{3mm}
%     \widehat{\Omega_{s_{k_1,...,k_l}}} = \prod_{1\leq i\leq N-1\atop {1\leq j\leq l\atop i\neq k_j}}\Omega_i,\hspace{3mm}%\\% \hspace{3mm},\hspace{3mm}
%     \\
%         &\widehat{d_{s_{k_1,...,k_l}}} = \prod_{1\leq i\leq N\atop {1\leq j\leq l\atop i\neq k_j}} d_i\\
%     \end{align*}
with the convention  that $\prod_{i\in \emptyset} X_i=1$. Similarly, we adopt the convention for the sum: $\sum_{i\in \emptyset}X_i=0$.
\paragraph{Transfer Matrix}
The use of transfer matrices to compute the propagation of rays in the sense of first order optics (small angles and spherical mirrors) is well known from opticians and makes computation easier. 
We refer the reader to \cite{PerezBook}  for more details on the construction of transfer matrices to model, at first order approximation, the propagation of light through different optical surfaces (lenses, mirrors, gratings) in homogeneous media.
Let us recall that, in this paper, we deal only with mirrors.
Let   $M_N =M_{V_{N}}M_{d_{N-1}}... M_{d_{1}}M_{V_1}$ be the transfer matrix of the system $[S_1,...,S_N]$ 
with
$
M_{V_k} = \left(\begin{array}{cc}
    1&0\\
    -V_k&1\\    
\end{array}\right)$
and $ M_{d_k} = \left(\begin{array}{cc}
    1&(-1)^kd_k\\
    0&1\\    
\end{array}\right)
     $
     where $V_k = 2(-1)^kc_k$ is the vergence of the $k$-th mirror in the local coordinate system associated to the optical axis,
     and $(-1)^kd_k>0$ the distance along the optical axis between  mirrors $S_{k}$ and $S_{k+1}$ (see \eqref{eq:courbures} and  \cref{fig:notationKorsh}).
     An incident ray $r_e = (x_e,\alpha_e)$ on mirror $S_1$ is transformed into an exit ray $r_s = (x_s,\alpha_s)$  on $S_N$ as  $r_s=M_Nr_e$.
     We have the following proposition.
\begin{prop}
    \label{prop:expressionMatrixTransfert}
    Let $N\geq 2$, the transfer matrix $M_N$ writes as
    \begin{equation}
        M_N=\frac{1}{\Omega_{s_{N-1}}}\left(\begin{array}{cc}\alpha_N&\beta_N\\ \gamma_N&\delta_N\end{array}\right)
        \label{eq:expressionM_N}
    \end{equation}
    where
   
    \begin{equation}
        \label{eq:abcd_Np_general}
        \begin{array}{cc}
            \alpha_N=\Omega_{s_{N-1}}^2+v_1\mathcal{S}_{N-1},\quad &
            \beta_N=\mathcal{S}_{N-1},\\
            \gamma_N= -\Omega_{s_{N-1}}^2(-1)^Nv_N'-\mathcal{S}_{N-1}v_1(-1)^Nv_N'+v_1,\quad&
            \delta_N=-\mathcal{S}_{N-1}(-1)^Nv_N'+1,
        \end{array}
    \end{equation}
   with   
        \begin{equation}
            \label{eq:defSNm1}
            \mathcal{S}_{N-1}=\sum_{l=1}^{N-1}(-1)^ld_l\Omega_l\prod_{l+1}^{N-1}\Omega_i^2.
        \end{equation}
\end{prop}
\begin{proof}
    Straightforward by induction.
\end{proof}
In the sequel, we consider only telescopes observing an object coming from infinity what leads to take $v_1=0$ in \eqref{eq:courbures} and \eqref{eq:abcd_Np_general}.
%We denote by $(e_1,e_2)$ the canonical basis of $\mathbb{R}^2$.

  \subsection{Polynomials systems for focal telescopes}
  \label{sec:polynomialFocalSystem}
  This subsection is dedicated to focal telescopes focusing  on a focal plane located at a finite distance $d_N$ w.r.t to $S_N$, what leads to $v_N'=\frac{1}{d_N}$ in \eqref{eq:courbures} and \eqref{eq:abcd_Np_general}. 
Let  $f\neq 0$ be the focal length of the telescope and $N$ be the number of mirrors composing it.
We give a polynomial description of the image, focal, Petzval and telecentricity constraints depending on $f$, $(\Omega_k)_{1\leq k\leq N-1}$ and $(d_k)_{1\leq k \leq N}$. 
\paragraph{Image constraint}
The total transfer matrix from the source to the focal plane is (see \eqref{eq:expressionM_N}) 
\[
M_{d_N}M_N = \frac{1}{\Omega_s}\left(\begin{array}{cc}\alpha_N+(-1)^Nd_N\gamma_N&\beta_N+\delta_N(-1)^Nd_N\\
    \gamma_N&\delta_N
\end{array}\right)=\left(\begin{array}{cc}a&b\\
c&d\end{array}\right)
\]
The image condition writes as $a=0$ 
what is  satisfied  thanks to \cref{prop:expressionMatrixTransfert} and recalling that $d_N=\frac{1}{v_N'}$.

\paragraph{Focal constraint}
The focal $f$ is the sensitivity of the lateral position $x_s$ on the focal plane w.r.t the entry angle $\alpha_e$.
By using notations of the above paragraph, the focal condition writes as $b=f$ and we denote this relation as $g_{1,N}=0$ with: 

\begin{equation}
    g_{1,N}=f \Omega_{s}- (-1)^Nd_N.
    \label{eq:C1focale}
\end{equation}
\paragraph{Petzval's constraint}
The vanishing Petzval curvature condition, introduced by Petzval in the mid-19th century enables to eliminate the field curvature, leading to a focal plane which is indeed plane.
 The Petzval constraint  writes  as the vanishing sum of curvatures along the optical axis that is $\sum_{k=1}^N(-1)^kc_k=0$. By mutiplying this latter condition  by the term $\Omega_s d_s$ which does not cancel over the constrained semi-algebraic set (see the next paragraph about the constraints), we obtain  $g_{2,N}=0$ with:
%  \begin{equation*}
%     \begin{aligned}\sum_{k=1}^N(-1)^kc_k=0&\Longleftrightarrow g_{2,N}=0
%     \end{aligned}
% \end{equation*}
% where $g_{2,N}$ is obtained by mutiplying \eqref{eq:courbures}  by $\Omega_s d_s$ which does not cancel over the constrained semi-algebraic set:
\begin{equation}
    g_{2,N}=\sum_{k=1}^{N-1}(-1)^{k+1}(1-\Omega_k)^2\widehat{\Omega_{s_k}}\widehat{d_{s_k}}+(-1)^Nd_{s_{N-1}}\Omega_s.
    \label{eq:C2Petzval}
\end{equation}

\paragraph{Telecentricity constraint}
The telecentricity condition is often used for spectro-imager telescopes. Indeed, for this kind of optical configurations, the source of the spectrometer is located at the focal plane of the imager. 
The entrance pupil   is imaged  at infinity by the imager so that the source of the spectrometer can be considered as ponctual.
%  each point corresponds to a field of the imager and with uniform   angular cones  parallel to each others.  
Assuming that the entrance pupil is located on the first mirror, and setting the exit pupil at a distance $z_N$ from the last mirror relatively to the optical axis, we obtain  the condition $\beta_N+\delta_Nz_N=0$.
% \begin{align}
%     %\left(0,\alpha_e\right)\underset{M_{z_N}M_N}{\longrightarrow}(0,\alpha_s),\ \forall \alpha_e
%     \exists \epsilon >0\ \forall \alpha_e\in  [0,\epsilon),\quad  \left\langle M_{z_N}M_N\left(\begin{array}{c}
%         0\\ \alpha_e
%     \end{array}\right),e_1\right\rangle=0
%     &\Longleftrightarrow \beta_N+z_N\delta_N=0
%   \label{eq:pupil}
% \end{align}
The telecentricity condition corresponds to the limit of this expression as 
 $z_N\to\infty$ which rewrites as $g_{3,N}=0$: 
% \begin{equation*}
%     \delta_N=0\Longleftrightarrow g_{3,N}=0
% \end{equation*}
% with 
\begin{equation}
    g_{3,N}=\delta_n=-\mathcal{S}_{N-1}+(-1)^Nd_N,
    \label{eq:C3TC}
\end{equation}
where we recall that $\mathcal{S}_{N-1}$ is defined in \eqref{eq:defSNm1}.
% \begin{equation}
%     \label{eq:defSigmaNm1}
%     \mathcal{S}_{k}=\sum_{l=1}^{k}(-1)^ld_l\Omega_l\prod_{i=l+1}^{k}\Omega_i^2.
% \end{equation}
\paragraph{Unknowns}
By homogeneity, without loss of generality, the focal $f$ can be taken equal to $\pm 1$.
Let $n\in \llbracket 1,3\rrbracket$ be the number of equations, the dimension of the affine space is $2N-1$.
 We define the polynomial sequence $\mathbf{g_{n,N}} =(g_{1,N},..,g_{n,N})\in \mathbb{Q}[X_1,\dots,X_{2N-1}]^n$
% \begin{equation}
%     \label{eq:defpolynomialsequenceg}
%     \mathbf{g_{n,N}} =(g_{1,N},..,g_{n,N})
% \end{equation}
where $(X_1,\dots,X_{2N-1})=(d_1,...,d_N,\Omega_1,...,\Omega_{N-1})$ 
and $t=2N-1-n$ the dimension of the affine variety $\mathbf{V}(\mathbf{g_{n,N}})$ (see \cref{sec:preliminaries}).
%we define  $S_{\{1,...,k\}}^{(N)}$ as the polynomial system 
%$$S_{\{1,...,k\}}^{(N)}=\{g_j\in \mathbb{Q}[\mathbf{x}],\ 1\leq j\leq k\}$$  
% where the indeterminate $(\mathbf{y},\mathbf{x})$ is 
% \begin{align*}
%     (\mathbf{y},\mathbf{x}) &= (d_1,...,d_N,\Omega_1,...,\Omega_{N-1})\in\mathbb{R}^{t+n}.\ \\
% \end{align*}
% The parameters $\mathbf{y}\in \mathbb{R}^t$ and the unknowns $\mathbf{x}\in \mathbb{R}^{n}$ will be precised according to the codimension of the considered system. 
%\paragraph{Summary.}
When $N$ is clearly specified, we denote by $\mathbf{g_n}$ instead of $\mathbf{g_{n,N}}$ and $g_{k}$ instead of $g_{k,N}$.% where the $g_k\in \mathbb{Q}[\mathbf{y},\mathbf{x}]$  are recalled here:
% \begin{equation}
%     \label{eq:summarizedEqFocal}
%     \begin{aligned}
%     g_1(\mathbf{y},\mathbf{x})&=f \Omega_{s}- (-1)^Nd_N\\
%     g_2(\mathbf{y},\mathbf{x})&= \sum_{k=1}^{N-1}(-1)^{k+1}(1-\Omega_k)^2\widehat{\Omega_{s_k}}\widehat{d_{s_k}}+(-1)^Nd_{s_{N-1}}\Omega_s=0\\
%     g_3(\mathbf{y},\mathbf{x})&=-\mathcal{S}_{N-1}+(-1)^Nd_N=0\\
%     \end{aligned}
% \end{equation}

\paragraph{Constraints\label{par:constraints_foc}}
The constraints correspond to  positiveness of the distances along the optical axis that is $(-1)^kd_k>0$ for all $k\in \llbracket 1,N\rrbracket$ and non null magnifications that is $\Omega_s\neq 0$.
Note that the focal condition \eqref{eq:C1focale}, combined with the requirement \( (-1)^N d_N > 0 \) and the fact that \( f = \pm 1 \), implies that \( \Omega_s \neq 0 \). Hence let $\mathcal{G}_o(X)$ be the logical semi-algebraic formula corresponding to these conditions, it is given by

% : 
% \begin{align*}
%      {\mathcal{G}_o}(X)=\land_{l=1}^{N}((-1)^ld_l>0) \land_{l=1}^{N-1}(\Omega_l\neq 0)\\
% \end{align*}
% Note that the focal condition \eqref{eq:C1focale}, combined with the requirement \( (-1)^N d_N > 0 \) and the fact that \( f = \pm 1 \), implies that \( \Omega_s \neq 0 \).
% Hence, ${\mathcal{G}_o}(\mathbf{y},\mathbf{x})$ can be reduced to 
\begin{align}
    \mathcal{G}_o(X)=\land_{l=1}^{N}((-1)^ld_l>0) \label{eq:constraintFocal}.
\end{align}

  \subsection{Polynomials systems for afocal telescope}
  \label{sec:polynomialAFocalSystem}
  This section is dedicated to afocal telescopes for which the locus of focused output rays lies at an infinite distance relative to \( S_N \), resulting in \( v_N' = 0 \) in \eqref{eq:courbures} and \eqref{eq:abcd_Np_general}.
 This condition can be understood as the limit of \eqref{eq:C1focale} as \( |f| \to \infty \). 
 In this case, it is straightforward to verify that \( \gamma_N = 0 \) in \eqref{eq:abcd_Np_general}.
 In this subsection, as $d_N$ is infinite, $d_{s}$ denotes the product $\prod_{1\leq j\leq N-1}d_j$ and $\widehat{d_{s_k}}=\frac{d_s}{d_k}$.
\paragraph{Magnification constraint}
The lateral magnification is  the sensitivity of the lateral exit position $x_s$ on the last mirror $S_N$ w.r.t the entrance position $x_e$ on the first mirror $S_1$.
Let $G$ be the  lateral magnification,  this condition writes as $\alpha_N=G$ and rewrites as  $h_{1,N}=0$
with
\begin{equation}
    h_{1,N}=\Omega_s-G.\label{eq:C1grandissement}
\end{equation}
\paragraph{Petzval  constraint}
As explained in \cref{sec:polynomialFocalSystem}, the Petzval condition writes as $h_{2,N}=0$ with
\begin{equation}
    \label{eq:C2PetzvalAfoc}
    h_{2,N}=\sum_{k=1}^{N-1}(-1)^{k}c_k=\sum_{k=1}^{N-1}(-1)^{k+1}(1-\Omega_k)^2\widehat{\Omega_{s_k}}\widehat{d_{s_k}}.
\end{equation}
%with $d_s=\prod_{1}^{N-1}d_k$ and $\widehat{d_{s_k}} = d_s/d_k$ for $k\in \{1,...,N-1\}$.
\paragraph{Pupil positions constraint}
Let \( z_0 \) and \( d_p \) denote the signed distances along the optical axis from the entrance pupil to the first mirror and from the exit pupil to the last mirror, respectively.
 The pupil writes as \( \alpha_N z_0 + \beta_N + d_p(\gamma_N z_0 + \delta_N) = 0 \),  rewritten as \( h_{3,N} = 0 \) with:
% \begin{align*}
%     %\left(0,\alpha_e\right)\underset{M_{d_p}M_NM_{z_0}}{\longrightarrow}(0,\alpha_s),\ \forall \alpha_e
%      \exists \epsilon >0\ \forall \alpha_e\in [0,\epsilon),\quad \left\langle M_{d_p}M_NM_{z_0}\left(\begin{array}{c}
%         0\\ \alpha_e
%     \end{array}\right),e_1\right\rangle=0
%     &\nonumber\Longleftrightarrow \alpha_Nz_0+\beta_N+d_p(\gamma z_0+\delta_N)=0\\
%     &\label{eq:C3PupAfoc}\Longleftrightarrow h_{3,N}=0
% \end{align*}
% with
\begin{equation}\label{eq:C3PupAfoc}
    h_{3,N}=\Omega_s^2z_0+\mathcal{S}_{N-1}+d_p.
\end{equation}
\paragraph{Unknowns}
By homogeneity of the equations, without loss of generality, we  set $d_1=-1$.
Let $n\in \llbracket1,3\rrbracket$ be the number of equations, the indeterminate $X$ is
\begin{align*}
   (X_1,\dots,X_{2N-2}) &=     (G,d_2,...,d_{N-1},\Omega_1,...,\Omega_{N-1})  \text{ if }n\in \{1,2\}\\
    (X_1,\dots,X_{2N}) &= (G,z_0,d_p,d_2,...,d_{N-1},\Omega_1,...,\Omega_{N-1}),\ \text{ if }n=3.
\end{align*}
Let  $\mathbf{h}_{n,N} =(h_{1,N},..,h_{n,N})\in \mathbb{Q}[X]^n$ be the polynomial sequence. 
Let $t=2N-n$ if $n\in \{1,2\}$ (resp. $t=2N-3$ if $n=3$) be the dimension of the affine variety $\mathbf{V}(\mathbf{h_{n,N}})$. 
% \begin{equation}
%     \label{eq:defpolynomialsequenceh}
%     \mathbf{h}_{n,N} =(h_{1,N},..,h_{n,N})\in \mathbb{Q}[\mathbf{y},\mathbf{x}]^n
% \end{equation}
% with
% %Let $1\leq k\leq 3$ and $S_{\{1,...,k\}}^{(N)}$ the polynomial system $\{g_j,\ 1\leq j\leq k\}$ considered over $\mathbb{Q}[x]$ with the indeterminate $x$ defined by
% \begin{align*}
%     (\mathbf{y},\mathbf{x}) &=     (G,d_2,...,d_{N-1},\Omega_1,...,\Omega_{N-1})  \text{ for }n\in \{1,2\}\\
%     (\mathbf{y},\mathbf{x}) &= (G,z_0,d_p,d_2,...,d_{N-1},\Omega_1,...,\Omega_{N-1}),\ \text{ for }n=3
% \end{align*}
% As for the focal case, the parameters $\mathbf{y}\in \mathbb{R}^t$ and the unknowns $\mathbf{x}\in \mathbb{R}^{n}$ will be precised according to the codimension of the considered system. 
When $N$  is clearly specified, we denote by $\mathbf{h_n}$ instead of $\mathbf{h_{n,N}}$ and $h_{k}$ instead of $h_{k,N}$.
%The equations considered in the afocal case for $N$ mirrors summarizes as the real vanishing of the following polynoms of $\mathbb{Q}[\mathbf{x}]$:
% \begin{equation}
%     \label{eq:summarizedEqAfocal}
%     \begin{aligned}
%     h_1(\mathbf{y},\mathbf{x})&= \Omega_{s}- G\\
%     h_2(\mathbf{y},\mathbf{x})&= \sum_{k=1}^{N-1}(-1)^{k+1}(1-\Omega_k)^2\widehat{\Omega_{s_k}}\widehat{d_{s_k}}=0\\
%     h_3(\mathbf{y},\mathbf{x})&=\Omega_s^2z_0+\mathcal{S}_{N-1}+d_p=0  \\
%     \end{aligned}
% \end{equation}
%As in the focal case, let be $\mathbf{g}_k$ the polynomial sequences  $\mathbf{g}_{k} =(g_1,..,g_k)\in \mathbb{Q}[\mathbf{x}]$ for $k\leq 3$, that we will rewrite as explained previously in  section~\ref{sec:prespolynomialFocalSystem}.
%As explained in section~\ref{sec:polynomialFocalSystem}, $\mathbf{h}_k$ we will be rewriten in an adhoc form which enables to characterize more easily its associated vanishing algebraic set. 
\paragraph{Constraints}
Similarly as explained in the corresponding subparagraph of \cref{sec:polynomialFocalSystem}, we define the set of constraints $\mathcal{G}_o(X)$ associated to the polynomial sequence $\mathbf{g_k}$ as:
\begin{align}
    \mathcal{G}_o(X) &= \land_{l=2}^{N-1}((-1)^ld_l>0)\land (G\neq 0)\ \label{eq:constraintAFocal}
\end{align}

As observed, the first-order equations \eqref{eq:C1focale}-\eqref{eq:C2Petzval}-\eqref{eq:C3TC} and \eqref{eq:C1grandissement}-\eqref{eq:C2PetzvalAfoc}-\eqref{eq:C3PupAfoc}, along with the constraints \eqref{eq:constraintFocal} and \eqref{eq:constraintAFocal}, 
are polynomial functions of the variables.
 As we will see in \cref{sec:optical_application}, additional constraints enforcing non-vanishing curvatures can be introduced to exclude configurations with planar mirrors,
  which will also be expressed as a non-zero polynomial condition. Therefore, as we are only interested in real solutions, we will apply tools from real algebraic geometry to study their topological properties.

  \section{Problem classification and real algebraic geometry}
  \label{sec:pb_classif}
  The study of semi-algebraic sets has a lot of applications, such as in robotics  \cite{Cox,CapcoEldin}, biology \cite{chauvin} or control theory \cite{HenrionMarx}.
  Combined with computer algebra, algorithms involved in real algebraic geometry enable to avoid numerical instabilities due to high non linearities \cite{NMRImagingClassif2003}.
  A direction of the research of this domain consists in designing new algorithms which enable to solve in finite time very important problems with a lot of applications such that computing at least one point per connected components of a semi-algebraic set \cite{schostElDin2003}, computing the dimension of semi-algebraic sets \cite{Vorobjov99,Lairez_2021}, deciding the connectivity between two points of a semi-algebraic set \cite{prebet2023computing} and computing a description of the algebraic set \cite{gaillard2024}.
  For example, in control theory, an important problem is to  characterize the region of controls that gives admissible solutions or to know if two admissible points can be linked by a continuous path in the admissible set. 
  % This questions are addressed theoretically in the field of real algebraic geometry as the problem of connectivity query, semi-algebraic connected components representant searching. 
  In this section, we present  the problem of classification that we address in this paper, and its formulation as the study of connected components of a semi-algebraic set.
  We take benefit from the special form of the polynomial sequences  in optics derived in  \cref{sec:polynomialFocalSystem} and \cref{sec:polynomialAFocalSystem} 
  by splitting the studied set in two parts separated by the locus of critical points of a well chosen canonical projection.
  Then, we show that each part is homeomorphic to its projection and we deduce the connected components of the initial set.
   Let us note that the  construction of the homeomorphisms  relies on a generic algorithm described in \cite{LE202225} which solves a root classification problem.
  Let us start by some preliminaries on real algebraic geometry.
  \subsection{Preliminaries}
  \label{sec:preliminaries}
  This section presents some  used concepts of real algebraic geometry which are used to solve polynomial equations under polynomial inequalities, the heart of this paper.
For more details on this we refer the reader to \cite{faugere:hal-01298887,basu2014algorithms,Cox}. 
\paragraph{Algebraic sets and ideals}
Let $d\in \mathbb{N}^\star$  and $\mathbb{F}$  be a sub-field of $\mathbb{C}$. 
Let $s\in \mathbb{N}^\star$ and $\mathbf{f}=(f_1,\dots,f_s)\in \mathbb{F}[\mathbf{x}]$ be a polynomial sequence with $\mathbf{x}=(x_1,\dots x_d)$. 
We denote by $\langle \mathbf{f}\rangle=\langle f_1,\dots,f_s\rangle$ the associated ideal generated by $\{f_1, \dots, f_s\}$  in the ring $A=\mathbb{F}[\mathbf{x}]$ defined by 
$\langle \mathbf{f}\rangle = \{g = \sum_{k=1}^sa_kf_k,\ a_k\in A\}$.
Let $I\subset\mathbb{F}[\mathbf{x}]$, the set
\[\mathbf{V}(I)=\{\mathbf{x}=(x_1,\dots,x_d)\in \mathbb{C}^d \ :\  \forall f\in I \quad f(\mathbf{x})=0\}
    \]
    is the algebraic set associated to $I$ i.e. the set of points in $\mathbb{C}^d$ at which all polynomials in $I$ vanish.
By abuse of notation we write $\mathbf{V}(\langle \mathbf{f}\rangle)=\mathbf{V}( \mathbf{f})$. 
Conversely, for an algebraic set $\mathcal{V}\subset \mathbb{C}^d$, we denote by 
\[
    \mathbf{I}(\mathcal{V})=\{p\in \mathbb{C}[\mathbf{X}]\ :\ \forall x\in \mathcal{V}\quad p(x)=0\}
    \]
    the radical ideal associated to $\mathcal{V}$. Let $I\subset \mathbb{C}[\mathbf{x}]$ be an ideal such that there exists an  algebraic set $\mathcal{V}\subset \mathbb{C}^d$ such that  $I = \mathbf{I}(\mathcal{V})$ then $\mathcal{V}=\mathbf{V}(I)$.
     However, in general $I(\mathbf{V}(I))\neq I$.
    More precisely, let $I\subset \mathbb{C}[\mathbf{x}]$ be an ideal, the Nulltstellensatz of Hilbert [\cite{faugere:hal-01298887}, [\cite{Cox}, Theorem 6, chapter 4, \S 1] states that  $\mathbf{I}(\mathbf{V}(I))=\sqrt{I}=\{f\in \mathbb{C}[\mathbf{x}],\exists k\in \mathbb{N}\ f^k\in I \}$ and conversely for $\mathbf{f}$  a sequence of polynoms in $\mathbb{C}[\mathbf{x}]$, the associated algebraic set verifies $\mathbf{V}(\sqrt{\langle \mathbf{f}\rangle} )=\mathbf{V}(\mathbf{f})$.
    The real trace  $\mathbf{V}(I) \cap \mathbb{R}^d$ is denoted $\mathbf{V}_{\mathbb{R}}(I)$.\\
    The dimension of an algebraic set $\mathcal{V}\subset\mathbb{C}^d$ is defined locally as $d-\textrm{rank}(\textrm{jac}(\mathbf{f}))$  and also as the largest number $r$ such that there exists $\{i_1,\dots,i_r\}\subset \{1,\dots,d\}$ such that the projection 
    $\pi : x\in \mathcal{V}\mapsto (x_{i_1},\dots.,x_{i_r})$ is surjective outside an affine variety $\mathcal{W}\subset \mathbb{C}^r$. The dimension of an ideal is the dimension of the associated algebraic set.
For $\mathcal{A}\subset\mathbb{C}^d$, we denote by $\overline{\mathcal{A}}$ the Zariski closure of $\mathcal{A}$ that is the smallest algebraic set containing containing $\mathcal{A}$.
An algebraic set $\mathcal{V}$ is irreducible if the following holds : $\mathcal{V}=\mathcal{V}_1\cup\mathcal{V}_2\Longrightarrow (\mathcal{V}=\mathcal{V}_1)\lor (\mathcal{V}=\mathcal{V}_2)$. The notion of irreducible algebraic sets is in one to one correspondence with the notion of prime ideals.
An algebraic set is equidimensional of dimension $t$ if it is the union of irreducible algebraic set of dimension $t$.\\
For an algebraic set $\mathcal{V}=\mathbf{V}(\mathbf{f})$ with $\langle \mathbf{f}\rangle$ radical, if $c$ is the co-dimension of $\mathcal{V}$, then the set of singular points of $\mathcal{V}$ is the set of points of $\mathcal{V}$ at which $\rank(\jac(\mathbf{f}))<c$ and it is denoted by $\sing(\mathcal{V})$. A smooth point of $\mathcal{V}$ is a non singular point of $\mathcal{V}$.
Let   $\pi : (x_1,\dots,x_d)\to (x_{l+1},\dots.,x_d)$, we call $\crit(\pi,\mathcal{V})$ the set of critical points of the restriction of $\pi$ to $\mathcal{V}$. 
If $c$ is the codimension of $\mathcal{V}$ and $\mathbf{f}=(f_1,\dots,f_s)$ generates the vanishing ideal associated to $\mathcal{V}$ then $\crit(\pi,\mathcal{V})$ is the set of smooth points of $\mathcal{V}$ where the Jacobian matrix associated to $ (f_1,\dots,f_s)$ w.r.t to $(x_1,\dots.,x_l)$  has rank less than $c$. When $s=c=l$ and $\mathcal{V}$ is smooth ($\sing(\mathcal{V})=\emptyset$) this set 
is the intersection of $\mathcal{V}$  with the hypersurface associated to the vanishing determinant of the Jacobian matrix of $(f_1,\dots,f_s)$ w.r.t to $(x_1,\dots.,x_l)$. We denote $\mathcal{K}(\pi,\mathcal{V})=\sing(\mathcal{V})\cup\crit(\pi,\mathcal{V})=\{\mathbf{x}\in \mathcal{V},\ \rank(\jac(\mathbf{f},(x_1,\dots,x_l)))<c \}$ which rewrites in the case $s=c=l$ as $\mathcal{K}(\pi,\mathcal{V})=\{\mathbf{x}\in\mathcal{V},\ \det(\jac(\mathbf{f},(x_1,\dots,x_l)))=0 \}$.

\paragraph{Semi-algebraic set}
A subset $E\subset \mathbb{R}^d$ is a semi-algebraic set if it is a finite union of basic constructible  sets of the form
$
    \{\mathbf{x}\in \mathbb{R}^d,\ \bigwedge_{P\in \mathcal{P}} (P(\mathbf{x})=0)\land \ \bigwedge_{Q\in \mathcal{Q}}(Q(\mathbf{x})>0)
    \}$
with $\mathcal{P},\ \mathcal{Q}$ subsets of $\mathbb{R}[\mathbf{x}]$.
In the sequel we assume that there exist  two polynomial sequences $\mathbf{f}=(f_1,\dots, f_s)\in \mathbb{R}[\mathbf{x}]^s$ and $\mathbf{g}=(g_1,\dots,g_{r})\in \mathbb{R}[\mathbf{x}]^r$ and a symbol sequence $\bm{\sigma}=(\sigma_1,\dots,\sigma_r)^r\in \{<,\leq,\neq\}^r$ with $s$ and $r$ two non-negative integers, such that 
\begin{equation}
\begin{aligned}
    E&=\{\mathbf{x}\in \mathbb{R}^d,\ f_1(\mathbf{x})=0,\dots,\ f_s(\mathbf{x})=0,\ g_1(\mathbf{x})\sigma_1 0,\dots, g_{r}(\mathbf{x}) \sigma_{r}0\}.
\end{aligned}
\label{eq:defE}
\end{equation}
Let us introduce some notations.
 For $\phi:\mathbb{R}^{d}\to \{0,1\}$ we denote by $\set{\phi}=\{\mathbf{x}\in\mathbb{R}^d\ :\ \phi(\mathbf{x})\}$.
In particular we denote   $\mathcal{G}=\mathbf{g} \bm{\sigma} 0=\bigwedge_1^rg_i\sigma_i0$ and $\set{\mathcal{G}}\subset\mathbb{R}^d$   its associated semi-algebraic set.
The set of real solutions satisfying $\mathbf{f}\bm{=}0$ is defined and denoted as
$\VR{\mathbf{f}} =\set{\mathbf{f}\bm{=}0}=\mathbf{V}(\mathbf{f})\cap \mathbb{R}^{d}$.
With these  notations the set $E$ given in \eqref{eq:defE} rewrites as $E = \VR{\mathbf{f}}\cap \set{\mathcal{G}}$.\\

\paragraph{Elimination theory}
 
\begin{theo}[The Elimination Theorem {[\cite{Cox}, chapter 3, \S 1\ ]}]
\label{theo:eliminationTheorem}
    Let $\mathbb{K}$ a field, $I\subset \mathbb{K}[x_1,\dots,x_d]$ be an ideal and let $G$ be a Groebner basis of $I$ w.r.t lexical order $x_1\succ x_2\succ \dots \succ x_d$. Then for every $0\leq l \leq d$, the set 
    \[G_l = G\cap \mathbb{K}[x_{l+1},\dots,x_d]
        \]
        is a Groebner basis of the $l$-elimination ideal $I_l = I\cap\mathbb{K}[x_{l+1},\dots,x_d]$.
\end{theo}
 
\begin{theo}[The Extension Theorem {[\cite{Cox}, chapter 3, \S 1]}]
    \label{theo:extensionTheorem}
    Let $I=\langle f_1,\dots, f_s\rangle\subset \mathbb{C}[x_1,\dots,x_d]$ and let $I_1$ the first elimination ideal of $I$.
    For each $1\leq i\leq s$, write $f_i$ in the form
    \[
        f_i = c_i(x_2,\dots,x_d)x_1^{N_i}+\text{ terms in which }x_1\text{ has degree }<N_i,
        \] 
        where $N_i\geq 0$ and $c_i\in \mathbb{C}[x_2,\dots,x_d]$ is nonzero. Suppose that we have a partial solution $(a_2,\dots,a_d)\in \mathbf{V}(I_1)$. 
        If $(a_2,\dots,a_n)\not\in\mathbf{V}(c_1,\dots,c_s)$, then there exists $a_1\in \mathbb{C}$ such that $(a_1,\dots,a_n)\in \mathbf{V}(I)$.
\end{theo}
\begin{theo}[The Closure Theorem {[\cite{Cox}, chapter 4, \S 7]}]
    \label{theo:closureTheorem}
Let $\mathcal{V}=\mathbf{V}(I)\subset \mathbb{C}^d$ and $d>l>0$, then 
\[\overline{\pi_l( \mathcal{V})}=\mathbf{V}(I_l)
\]
and there exists an affine variety $\mathcal{W}\subset \mathbf{V}(I_l)$ such that 
\[
    \mathbf{V}(I_l)\backslash \mathcal{W}\subset \pi_l(\mathcal{V})\text{ and }\overline{\mathbf{V}(I_l)\backslash \mathcal{W}}=\mathbf{V}(I_l)
    \]
    where the closure is taken in the Zariski sense [\cite{Cox}, chapter 4, \S 4] and $\pi_l(x_1,\dots,x_d)= (x_{l+1},\dots,x_d)$.

    \end{theo}
    The Elimination Theorem and the Closure Theorem gives an algorithm to compute the Zariski closure of projection of algebraic sets. Indeed, it suffices to compute a Groebner basis of $I(\mathcal{V})$ and to keep only the elements of $G$ which belong to $\mathbb{C}[x_{l+1},\dots,x_d]$.
    Namely,    the closure theorem, in some way, precises the extension theorem in the following sense :   we can extend a  solution of $\mathbf{b}\in \mathbf{V}(I_l)$ to a solution in $\mathbf{V}(I)$ almost everywhere in the  Zariski sense.
    
Besides, a consequence is that the surjectivity of $\pi_l : \mathcal{V}\to\mathbb{C}^{d-l}$ up to a sub variety $\mathcal{W}\subset \mathbb{C}^{d-l}$ is reached when $\mathbf{V}({I}_l)=\mathbb{C}^{d-l}$ or equivalently when ${I}_l=\{0\}$. 
\paragraph{Classification and connected components characterization problems}
For a topological space $X$ we note by $\cc{X}$ the set of its path connected components.
Let  $E\subset\mathbb{R}^d$ be a semi-algebraic set (see \eqref{eq:defE}).
\noindent We formulate the problem of classification of solutions lying in $E$ as the study of  $\cc{E}$ or said differently  the set of equivalent classes in $E$ in the sense of homotopy: 
\begin{align}
    \label{eq:homotopyEquivalence}
    \mathbf{x}\overset{E}{\sim}\mathbf{y}&\Longleftrightarrow \exists \gamma\in C^{0}([0,1],E),\ \gamma(0)=\mathbf{x},\  \text{and } \gamma(1)=\mathbf{y}
    % &\overset{\text{def}}{\Longleftrightarrow}\exists C\in C\!\!\!C(E_{admissible}),\ x\in C\ \text{ and } y\in C
\end{align}

\paragraph{Real root classification algorithm}
Let $t,n\in \mathbb{N}^\star$ such that $d = t+n$, in view to characterize the different connected components of $E$, we propose to firstly classify the roots of  $\mathbf{V}_{\mathbb{R}}(\mathbf{f}) = \mathbf{V}(\mathbf{f})\cap \mathbb{R}^{t+n}$ in the parameters space as defined in the following.
 Let $I=\langle \mathbf{f}\rangle\subset\mathbb{Q}[\mathbf{y},\mathbf{x}]$ with $\mathbf{y}=(y_1,\dots,y_t)$ and $\mathbf{x}=(x_1,\dots,x_n)$. We name $\mathbf{y}$ as the parameters and $\mathbf{x}$ as the unknowns. We consider a monomial order $\mathcal{M}$ such that $\mathcal{M}(\mathbf{x})\succ\mathcal{M}(\mathbf{y})$.
We assume that the  $n$-th elimination ideal relatively to $\mathcal{M}$ is $I_n = I\cap \mathbb{Q}[\mathbf{y}] =\{0\}$ or equivalently $\mathbf{V}(I_n)=\mathbb{C}^{t}$.
 According to the extension theorem, this last hypothesis makes the projection $\pi:(\mathbf{y},\mathbf{x})\mapsto \mathbf{y}$  surjective from $\mathbf{V}(I)$ into a Zariski open set $\mathcal{O}\subset\mathbb{C}^{t}$.
  In the sequel, for $\bm{\eta}\in \mathbb{C}^t$, we denote by $\varphi_\eta:f\mapsto f(\bm{\eta},.)$ the specialization map from $\mathbb{C}(\mathbf{y})[\mathbf{x}]\to \mathbb{C}[\mathbf{x}]$.
 Let us state  the main result of \cite{LE202225}. Let us assume that $\langle\mathbf{f}\rangle$ is radical and $\mathcal{V}=\mathbf{V}(\mathbf{f})$ satisfies 
 \begin{hyp}
    \label{hyp:finiteFiber}
    There is a Zariski open set $\mathcal{O}\subset \mathbb{C}^{t}$ such that for all $\mathbf{y}\in\mathbb{C}^{t}$, $\pi^{-1}(\mathbf{y})\cap \mathcal{V}$ is finite.
 \end{hyp}
 \noindent The algorithm given in \cite{LE202225} aims to solve this  root classification problem :
\begin{pb}
    \label{pb:classif}
\begin{itemize}
    \item Input: $\mathbf{f}$ such that $\mathcal{V}=\mathbf{V}(\mathbf{f})$ satisfies \cref{hyp:finiteFiber}, 
    \item Output: a collection of semi-algebraic sets $S_1,\dots,S_m$ such that
    \begin{itemize}
        \item[(i)] The number of real solutions in $\pi^{-1}(\mathbf{y})\cap \mathcal{V}$ is constant on $S_i$, $1\leq i\leq m$,
        \item[(ii)] The union of $S_i$'s is dense in $\mathbb{R}^{t}$
    \end{itemize}
\end{itemize}
The $S_i$ will be described by $(\Phi_i,\mathbf{y}_i,r_i)$ with $\Phi_i$ a semi-algebraic formula describing $S_i$, $\mathbf{y}_i$ a sampling point in $\mathbb{Q}^t$ and $r_i$ the corresponding number of real solutions. 
\end{pb}
\begin{re} 
\label{re:projectionEqualQuantifierElimination}
Let  $\Phi(\mathbf{y})=\lor_{i=1}^k \phi_i(\mathbf{y})$ be the union of the semi-algebraic formula solving \cref{pb:classif}, then  $Z(\Phi)$ is dense in $\pi(\mathbf{V}_\mathbb{R}(\mathbf{f}))$.
%  which is given by  $\mathcal{Z}(\overline{\Phi})$ where $\overline{\Phi} $
%  is solution of the following quantifier  elimination problem : 
%  \[ 
%     \exists \mathbf{x}\in R^{n},\  \Psi(\mathbf{y},\mathbf{x})\Leftrightarrow \overline{\Phi}(\mathbf{y})
%     \]
%     for $\mathbf{y}\in \mathbb{R}^{t}$ fixed and where $\Psi(\mathbf{y},\mathbf{x})=(\mathbf{y},\mathbf{x})\in \mathbf{V}_\mathbb{R}(\mathbf{f}(\mathbf{y},.))$ with $\mathbf{f}(\mathbf{y},.)=\varphi_\mathbf{y}(\mathbf{f})$  the specialization of $\mathbf{f}$ at $\mathbf{y}$.

\end{re}
In \cite{LE202225}, authors solve this problem generically through the use of Hermite matrices to deduce via their signature the semi-algebraic representation of the $S_i$. 
%We summarize here the key points of this algorithm to be self-content and refer the reader to \cite{LE202225}  for more details.
Namely, let  $G$ be a Groebner basis of $I$ with $\mathcal{M}(\mathbf{x})\succ\mathcal{M}(\mathbf{y})$ with $\mathcal{M}$ a monomial order\footnote{In \cite{LE202225} authors takes $\mathcal{M}=\textrm{grevlex}$ for comptational complexity reasons but the theory holds for whatever monomial order $\mathcal{M}$.}, and $w_\infty$ be the square-free part of $\prod_{g\in G}\textrm{lc}_x(g)$ and $\mathcal{W}_\infty=\mathbf{V}(w_\infty)\subset\mathbb{C}^t$. 
Considering $\mathbf{f}$ over $\mathbb{K}[\mathbf{x}]$ with $\mathbb{K}=\mathbb{Q}(\mathbf{y})$ enables to show that  $\langle \mathbf{f}\rangle_{\mathbb{K}}$ is zero dimensional and so the quotient ring $A_\mathbb{K}=\mathbb{K}[\mathbf{x}]\backslash \langle \mathbf{f}\rangle_{\mathbb{K}}$ is a finite dimensional $\mathbb{K}$-vector space. 
Let  $\delta$ be its dimension and $\mathcal{B}=(b_1,\dots,b_\delta)$ its basis. At this step, the notion of parametric Hermite matrix [\cite{basubook}-4.6] is introduced and defined as the matrix representation of the quadratic form of $A_\mathbb{K}\times A_\mathbb{K}$ associated to $(f,g)\mapsto \textrm{tr}(\mathcal{L}_{fg})$ where $\mathcal{L}_f$ for $f\in A_{\mathbb{K}}$ is the multiplication endomorphism of $A_\mathbb{K}$.
 On the basis $\mathcal{B}$, this quadratic form can be represented by a matrix $\mathcal{H} = (h_{i,j})_{1\leq i,j\leq \delta}$ where $h_{i,j}=\textrm{tr}(\mathcal{L}_{b_ib_j})$, whose the entries are lying in $\mathbb{K}$.
Authors carefully makes the link between the parametric Hermite matrix $\mathcal{H}(\bm{\eta})$ and the usual Hermite matrix $\mathcal{H}_{\bm{\eta}}$ associated to the  specialized ideal $I_{\bm\eta}=\langle \varphi_{\bm{\eta}}(\mathbf{f})\rangle$ at some $\bm{\eta}\in \mathbb{C}^t\backslash \mathcal{W}_\infty$ by showing that $\mathcal{H}(\bm{\eta})=\mathcal{H}_{\bm{\eta}}$. 
This enables to use well known results on Hermite matrices associated to the zero dimensional ideal $I_{\bm{\eta}}$, that is, the $\rank$ of $\mathcal{H}({\bm{\eta}})$ is equal to the number of distinct complex roots and its $\signat$ to the number of distinct real roots of $\mathbf{f}(\bm{\eta},.)$ [\cite{basubook}-Theorem~4.102].
 Finally, the sequence $W=[M_1,\dots,M_\delta]$ of the leading minors of $\mathcal{H}$, and $w_\mathcal{H}=\mathbf{n}/\textrm{gcd}(\mathbf{n},w_\infty)$ where $\mathbf{n}$ is the square-free product of $\det(\mathcal{H})$ is introduced. 
 Let  $\mathcal{W}_\mathcal{H}=\mathbf{V}(w_\mathcal{H})\subset\mathbb{C}^t$ be the vanishing algebraic set associated to $w_\mathcal{H}$.
Let us defined the sign function   $\sign$ as $\sign(x)=-1$ if $x<0$, $\sign(x)=+1$ if $x>0$ and $\sign(x)=0$ if $x=0$.
The semi algebraic cells $(S_i)_i$ of \cref{pb:classif} deduce from the following representation  
\begin{equation}
    \label{alg:rootClassif}
\begin{aligned}
\Phi(\mathbf{y})=\bigvee_{\bm{\eta}\in L}\phi_{\bm{\eta}}(\mathbf{y})
\text{ with } \quad \phi_{\bm{\eta}}(\mathbf{y})=\left(\bigwedge_{k=1}^\delta[\sign(M_k(\mathbf{y}))=\sign(M_k(\bm{\eta}))]\right)\bigwedge (w_\infty(\mathbf{y})\neq 0)
\end{aligned}
\end{equation}
for $\bm{\eta}$ lying in a set $L\subset\mathbb{Q}^t$ sampling the connected components of $\mathbb{R}^t\backslash(\mathcal{W}_\mathcal{H}\cup \mathcal{W}_\mathcal{\infty})$ \cite{Collins:75,schostElDin2003}. 
 %For example $L$ can be the output of a Cylindrical Decomposition Algorithm \cite{Collins:75} or of more recent Algorithms based on Morse Theory \cite{schostElDin2003}\footnote{ Let us note that the complexity of Cylindrical Decomposition Algorithm is doubly exponential w.r.t the number of unknowns while the other cited algorithm is simply exponential.}. 
 Note that if $\bm{\eta} \in \mathbb{C}^t$ is such that $\signat(\mathcal{H}(\eta)) = 0$, this implies that the system $\mathbf{f}(\eta, .)$ has no real solutions. Hence classifying the real roots of $\mathbf{f}(\eta,.)$  leads to consider only the subset $L_o = \{\eta\in L,\ \signat(\mathcal{H}(\eta))\neq 0\}$. 
 %We denote by $L_{oo}=\textrm{EliminateRedundantSign}(L_o)$ the procedure eliminating redundant elements in $\eta\in L_o$ leading to same sequence  $\sign(W(\eta))$  that is $L_{oo}\subset L_o$ is such that $\{\sign(W(\eta)),\ \eta\in L_{oo}\}=\{\sign(W(\eta)), \eta\in L_o\}$ and for all $(\eta,\eta')\in L_{oo}^2$ such that $\eta\neq\eta'$, $\sign (W(\eta))\neq \sign(W(\eta'))$. The real root classification algorithm is given in Algorithm~\ref{alg:rootClassif}.
% \begin{algorithm}
% \caption{Parametric Hermite matrix based root classification algorithm \cite{LE202225}}\label{alg:rootClassif}
%  \begin{algorithmic}
% \STATE $L\longleftarrow \textrm{Sample}(w_hw_\infty\neq 0)$
% \STATE Compute $L_o = \{\eta\in L,\ \sign(\mathcal{H}(\eta)\neq 0\}$
% %\State  $L_{oo}\longleftarrow \textrm{EliminateRedundantSign}(L_o)$
% \STATE $\Phi(\mathbf{y})=\bullet$
% \FOR{$\eta\in L_{o}$} 
%     \STATE Compute $W_\eta=[M_1(\eta),\dots,M_\delta(\eta)]$
%     \STATE Compute $\phi_\eta(\mathbf{y})=(\land_{k=1}^\delta\sign(M_k(\mathbf{y}))=\sign(M_k(\eta)))\land (w_\infty\neq 0)$
%     \STATE $\Phi(\mathbf{y})\longleftarrow \{\Phi(\mathbf{y}),\ \phi_\eta(\mathbf{y})\}$
% \ENDFOR 
% \STATE Return $\Phi$, $L_{o}$
% \STATE 

%  \end{algorithmic}
% \end{algorithm}

\begin{lemma}[\cite{LE202225}-Prop. 11]
\label{lemma:linkWhCritvalue}
    Grant  Assumption~\ref{hyp:finiteFiber},  $\pi(K(\pi,\mathcal{V}))\cup \mathcal{W}_\infty\subset \mathcal{W}_\mathcal{H}\cup \mathcal{W}_\infty$.
\end{lemma}
\cref{lemma:linkWhCritvalue} shows that the algebraic set $\mathcal{W}_\mathcal{H}$   is intimately linked to $\mathcal{K}(\pi,\mathcal{V})$. 
Besides, combined with the implicit theorem, the lemma provides  that on each $S_i$ there is a constant number $r_i$ of continuous function on the connected components of $S_i$ called \textit{branch} solution and denoted $\xi_{1,S_i},\dots,\xi_{r_i,S_i}$ from $S_i$ to $\mathcal{V}\cap\mathbb{R}^d$.
 Let  $(C_{i,j}')_j$ be the connected components of $S_i$ and $\xi_{1,C_{i,j}'},\dots,\xi_{r_i,C_{i,j}'}$ such that $\xi_{k,C_{i,j}'}={\xi_{k,S_i}}_{|C_{i,j}'}$. 

\paragraph{Inequalities constraints and branch extension}
  We set $Q\in \mathbb{Q}[\mathbf{y},\mathbf{x}]$, the square-free part of the product $g_1\dots g_r$. 
 Let us assume that  $\mathcal{V}_Q=\mathbf{V}(\langle \mathbf{f}\rangle +\langle Q\rangle)$  is of dimension $t-1$ and its projection is included in the vanishing algebraic set associated to the polynomial $w_Q\in \mathbb{Q}[\mathbf{y}]$ and denoted by $\mathcal{W}_Q =\mathbf{V}(w_Q)$, that is $\overline{\pi(\mathcal{V}_Q)}\subset\mathcal{W}_Q$. 
 We denote by $w_B=w_{\mathcal{H}}w_\infty w_Q$ the so-called border polynomial \cite{YangXia} and $\mathcal{W}_B$ its vanishing algebraic set. 
Applying the algorithm of \cite{LE202225} gives the sequence of sets $(S_i)_{1\leq i\leq m}$ such that on each $S_i$  the number of real roots is constant.
 We define $F_o =\overline{\cup_{i=1}^mS_i}=\overline{\pi(\mathbf{V}_\mathbb{R}(\mathbf{f}))}$ and recall that $E=\mathbf{V}_\mathbb{R}(\mathbf{f})\cap \set{\mathcal{G}}$.\\
\begin{lemma}
    \label{le:branchSolutionContinuOverComplementaryW_HW_infty}
    Let  $(S_i,r_i)_{1\leq i\leq m}$ be the sequence of semi-algebraic sets and  number of real solutions, outputs of \cref{pb:classif}.
    For each $C\in \cc{F_o\backslash \mathcal{W}_B}$  there exists  $i\in \{1,\dots,m\}$, $J\subset\{1,\dots,r_i\}$ and  a sequence of functions $(\xi_{j})_{j\in J}$, such that $S_i\supset C$  and $\xi_j\in C^0(C,E)$ for all $j\in J$.
\end{lemma}

\begin{proof}
    Let $C\in \cc{F_o\backslash \mathcal{W}_B}$, there exists $i\in \{1,\dots,m\}$ such that $C$ is included in a certain $S_i$ itself included in $F_o\backslash (\mathcal{W}_\infty\cup\mathcal{W}_H)$.
     Thanks to \cref{lemma:linkWhCritvalue}, for any $\bm{\eta}\in C$, $\pi^{-1}(\bm{\eta})$ does not meet $\mathcal{K}(\pi,\mathcal{V})$ what enables to show, thanks to the implicit function theorem,  the existence of $r_{i}$ continuous functions from $C$ to $\mathbf{V}_\mathbb{R}(\mathbf{f})$ denoted $\xi_j$ for $1\leq j\leq r_i$. 
    Since $C\subset\mathbb{R}^t\backslash \mathcal{W}_Q$ and $\mathcal{W}_Q\supset\pi(\mathcal{V}_Q)$ it follows that $\xi_j(\bm{\eta})\in \mathbf{V}_\mathbb{R}(\mathbf{f})\backslash \partial E$. 
    Thus, either $\xi_j(C)\subset \overset{\circ}{E}$ or $\xi_j(C)\subset {^cE}$ where $\overset{\circ}{E}$ and $ {^cE}$ represent the interior and the complement of $E$ as subset of $\mathbf{V}_\mathbb{R}(\mathbf{f})$,
relatively to the Euclidean topology.
     In the first case the $\xi_j(C)$ is retained, while in the second case, $\xi_j(C)$ is excluded.
     At the end, we obtain a set $J\subset\{1,\dots,r_i\}$ such that for each $j\in J$, $\xi_j$ is continuous from $C$ to $E$ what completes the proof.
\end{proof}
Let us note that the obtained connected cells defined as the image of a certain connected component of $F_o\backslash\mathcal{W}_B$ by $\xi_j$ is not necessarely a connected components of $E$. 
Firstly because a point $\eta\in \mathcal{W}_Q$ is not necessarely extendable to a solution $(\bm{\eta},\bm{\chi})\in \mathbf{V}(\mathbf{f})$ (see \cref{theo:extensionTheorem}), and
 a special consideration must be done to the sets $\mathcal{W}_\infty$ and $\mathcal{W}_\mathcal{H}$ on  which some continuous connexions in $E$ can be done as we will see in a special triangular quadratic case studied in the following subsection.

  \subsection{Special case of quadratic triangular system }
  \label{sec:triangular_system}
  
In the sequel of this subsection we do some assumptions which are verified by equations and  constraints derived in \cref{sec:polynomialFocalSystem} and~\cref{sec:polynomialAFocalSystem}.
\paragraph{Notations and statement of the problem} 
Let us recall that $\pi$ is the canonical projection $\pi:\mathbb{R}^{t+n}\ni(\mathbf{y},\mathbf{x})\mapsto \mathbf{y}\in\mathbb{R}^t$.
We denote by $(Q_k)_{k\in \mathcal{J}}$ the irreducible factors of $\mathcal{Q}$ for some  $\mathcal{J}\subset\mathbb{N}$ finite and $\mathcal{G}$ the logical clause associated to  inequalities.
%Let us recall that $\mathcal{G}=\mathbf{g}\mathbf{\sigma}0=\land_{k=1}^rg_k\sigma_k 0$ .
Let $F\subset\mathbb{R}^t$ and $\xi:F\subset\mathbb{R}^{t}\to \mathbb{R}^{n}$, we define $\Gamma(\xi)=\{(\mathbf{y},\xi(\mathbf{y})),\ \mathbf{y}\in F\}\subset\mathbb{R}^{t}\times\mathbb{R}^{n}$ the graph of $\xi$.
\begin{hyp}\label{hyp:codim}
    We assume  that $n\in \{2,3\}$ and there exists  $\mathbf{f}\in \mathbb{Q}[\mathbf{y},\mathbf{x}]^n$ such that
    \begin{itemize}
        \item[(a)]     $E=\mathbf{V}_{\mathbb{R}}(\f)\cap\set{\mathcal{G}}$,

    \item[(b)]    
There exists $\mathbb{R}^{t+n}\supset E_o\supset E$ and $\ftilde\in\mathbb{Q}[\mathbf{y},\mathbf{x}]^{n+1}$ such that  $E_o\cap \mathbf{V}(\f) =E_o\cap \mathbf{V}(\ftilde)$ where $\ftilde$ is  given by
\begin{equation}
\ftilde(\mathbf{y},\mathbf{x})=
\left(\begin{array}{c}
         A_1x_1^2+B_1x_1+C_1 \\
         A_2x_2+B_2\\
         \vdots\\
         A_{n}x_n+B_n\\
         x_nx_{n-1}-C_0
    \end{array}\right)=\left(\begin{array}{c}\tilde{f_1}\\ \vdots\\ \tilde{f}_{n+1}\end{array}\right)
    \label{eq:triangularQuadraticSystemCodim23}
    \end{equation}
    with $(A_1,\dots,A_n,B_1,C_1,C_0)\in \mathbb{Q}[\mathbf{y}]^{n+3}$ and 
    $B_k=u_kx_1+v_k$ with $(u_k,v_k)\in \mathbb{Q}[\mathbf{y}]^2$ for $k\in J_n=\{2,\dots, n \}$.
    \item[(c)]    \begin{align*}
        \gcd(A_1,B_1,C_1)&=1\\
        \forall k\in J_n\quad \gcd(A_k,u_k,v_k)&=1\\  
    \end{align*}
    \[
        \begin{array}{c}
            \exists \n\in \mathbb{Q}[\mathbf{y}]\ 
                \ : \ \forall k\in J_n\\ 
            \begin{aligned}                   
                        \gcd(\n,A_k)=1                 
                &\text{ and}&
                \left\lbrace
                \begin{aligned}
                            A_2C_0+B_2x_1&=\n \tilde{f_1}
                            \text{ i.e. }
                    \left\lbrace \begin{array}{c}
                        u_2=A_1\n \\
                        v_2=B_1\n\\
                        A_2C_0=C_1\n\\
                    \end{array} \right.
                    \hquad &\text{ if } n=2,\\
                    A_2A_3C_0-B_2B_3&=\n \tilde{f_1}
                    \text{ i.e. }
                        \left\lbrace \begin{array}{c}
                        u_2u_3=A_1\n \\
                        v_3u_2+u_3v_2=B_1\n\\
                        v_3v_2-A_2A_3C_0=C_1\n\\
                    \end{array} \right.
                    \hquad &\text{ if }n=3.\\
                \end{aligned}
                \right.
            \end{aligned}
        \end{array}
    \]
   
\end{itemize}
\end{hyp}
\begin{re}\label{re:whyftilde}
    For $n=2$, \cref{hyp:codim} could be replaced by $\mathbf{f}=(A_1x_1^2+B_1x_1+C_1,x_1x_2-C_0)$ and  $\{\tilde{f_1},\tilde{f_2},\tilde{f_3}\}$ is a Groebner basis of $\langle f_1,f_2\rangle$ for $\texttt{revlex}(\mathbf{x})\succ\texttt{revlex}(\mathbf{y})$.
    The shape of $\ftilde$ is convenient  to describe  the definition domain of the map solution (see \eqref{eq:defxieps}).
    Let us note that the $n+1$-th equation is redundant with the $n$ first equations:
    \begin{align*}
        &x_1\tilde{f_2}-A_2\tilde{f_3}=\n \tilde{f_1}\quad \text{ if }n=2,\\ 
            &\tilde{f_2}\tilde{f_3}-A_2A_3\tilde{f_4}-B_2\tilde{f_3}-B_3\tilde{f_2}=\n \tilde{f_1}\quad \text{ if }n=3,
    \end{align*}
     but it is needed to ensure the finiteness of the fiber $\pi^{-1}(\eta)$ for $\eta$ lying in  $\cup_{1\leq k\leq n}\mathbf{V}_\mathbb{R}(A_k))$ as we will se latter.
\end{re}
% Let  $G$ be a Groebner basis of $\langle {\ftilde}\rangle$ for a given monomial order $\mathcal{M}$ such that $\mathcal{M}(\mathbf{x})\succ\mathcal{M}(\mathbf{y})$, then following notations of the previous paragraph,
% the root classification \cref{alg:rootClassif} gives that there exists two real  continuous function solution of $\ftilde(\mathbf{y},.)=0$ on $F_o\backslash\mathcal{W}_\infty$ where $F_o=\set{\Delta\geq 0}\subset\mathbb{R}^t$ with  $\mathcal{W}_\infty$ the vanishing algebraic set associated to  $w_\infty=\textrm{sqf}(\prod_{g\in G}\textrm{lc}_x(g))$\footnote{$\textrm{sqf}$ stands for square free part of the product.} and $\Delta=w_\mathcal{H}=B_1^2-4A_1C_1$. 
% As explained in \cref{re:whyftilde}, we don't use Groebner basis here because \eqref{eq:triangularQuadraticSystemCodim23} has the advantage to define a solution map whose the definition domain is more easy to characterize.
Let $\Delta=w_\mathcal{H}=B_1^2-4A_1C_1$, $\mathcal{W}_{\mathcal{H}}=\mathbf{V}_\mathbb{R}(w_\mathcal{H})$ and $F_o=\set{\Delta\geq 0}\subset\mathbb{R}^t$,
similarly as  done by the classification \cref{alg:rootClassif}, the number of real root of $\ftilde$  is constant over each connected components of $\mathbb{R}^2\backslash \left(\mathcal{W}_\infty\cup\mathcal{W}_{\mathcal{H}}\right)$ where $\mathcal{W}_\infty = \mathbf{V}_\mathbb{R}(w_\infty)$ with $w_\infty=\textrm{sqf}(\prod_{k=1}^nA_k)$ where $\textrm{sqf}(P)$ is the square free part of $P\in \mathbb{R}[\mathbf{y},\mathbf{x}]$.
%In parallel, we denote by $\wt{w_\infty}$ the square free part  of the product of the $A_k$ that is $\wt{w_\infty}=\textrm{sqf}(\prod_{k=1}^nA_k)$ and
 Let $\epsilon\in \{-1,1\}$, we set 
\begin{equation}\label{eq:defx1eps}
x_1^{(\epsilon)} =\frac{-B_1+\epsilon\sqrt{\Delta}}{2A_1},\quad x_k^{(\epsilon)}=-\frac{B_k^{(\epsilon)}}{A_k}\quad \text{ where } B_k^{(\epsilon)}=\varphi^{(\epsilon)}(B_k)\quad  \forall k\in J_n,
\end{equation}
with $\varphi^{(\epsilon)}:\mathbb{Q}[\mathbf{y},x_1]\to C^0(F_o\backslash \mathcal{W}_\infty) $ the specialisation map such that $\varphi^{(\epsilon)}(p)=p(.,x_1^{(\epsilon)}(.))$
and let us define  the branch solution
\begin{equation} \label{eq:defxieps}
    \xi^{(\epsilon)}:\mathbf{y}\mapsto (x_1^{(\epsilon)}(\mathbf{y}),...,x_n^{(\epsilon)}(\mathbf{y}))
\end{equation}
 which is continuously defined on $F_o\backslash\mathcal{W}_\infty$ and where we  abuse  notation by writing $\mathcal{W}_\infty$ instead of $\mathcal{W}_\infty\cap F_o$.
%The expression of the $x_k$ shows that $\mathcal{W}_\infty\subset\wt{\mathcal{W}_\infty}$.
As root of polynomial system, $\xi^{(\epsilon)}$ is a semi-algebraic function whose the graph co-restricted to $E$ verifies
\[
 \Gamma(\xi^{(\epsilon)})\cap E\subset\{(\mathbf{y},\mathbf{x})\in \mathbb{R}^{n+t}\ : \ \mathbf{f}(\mathbf{y},\mathbf{x})=0\land \epsilon (2A_1x_1+B_1)\geq  0\land \mathcal{G}(\mathbf{y},\mathbf{x})\}.
\]
Let us note that the inclusion can be strict, particularly if  the fiber $\pi^{-1}(\mathbf{\eta})$ at points $\mathbf{\eta}$ lying in $\mathcal{W}_\infty$ is not finite.
%$E\cap\mathbf{V}(\langle \mathbf{f}\rangle +\langle w_\infty \rangle)$.
The sequel will be mainly devoted to show that under some asssumptions verified for our optical application, the equality holds true.
\paragraph{Connected components study of $E$}

The algebraic consequence of \cref{hyp:codim} is 
\begin{lemma}
    \label{le:compatibility}
    Grant \cref{hyp:codim}, there exists $(q_k,p_k)\in \mathbb{Q}[\mathbf{y}]^2 $ such that for all $k\in J_n$: 
    \begin{equation}
        \label{eq:PkQkLienDelta}
        \left\lbrace
        \begin{aligned}
            2A_1v_k-B_1u_k&=u_kp_k\\
            \Delta&=p_k^2+q_{k}A_k
        \end{aligned}
        \right.
    \end{equation}
    with 
\begin{subequations}
    \begin{equation}
        \label{eq:p_kcodim2}
        p_2=\frac{v_2}{\n}=B_1,\quad q_{2}=-\frac{4A_1C_0}{\n} \quad\text{ if }n=2
    \end{equation}
    \begin{equation}
        \label{eq:p_kcodim3}
        p_2 = -p_3=\frac{(u_3v_2-u_2v_3)}{\n}, \quad q_{2}=\frac{4A_3A_1C_0}{\n},\quad  q_{3}=\frac{4A_2A_1C_0}{\n}\quad\text{ if }n=3.
    \end{equation}
\end{subequations}       
\end{lemma}
\begin{proof}
The following relations are easily obtained:
 \[\left\lbrace
        \begin{aligned}
        2A_1v_2-B_1u_2&=u_2p_2\\ \Delta&=p_2^2+q_2A_2
        \end{aligned}\right.\]
        with    
        \[
        \left\lbrace \begin{aligned}
            p_2&=\frac{v_2}{\n}=B_1\in\mathbb{Q}[\mathbf{y}]\\
            q_2&=-\frac{4A_1C_0}{\n}
        \end{aligned}\right.\text{ if }n=2,
        \quad \left\lbrace \begin{aligned}
            p_2&=\frac{u_3v_2-u_2v_3}{\n}\\
            q_2&=-\frac{4A_1A_3C_0}{\n}
        \end{aligned}\right.\text{ if }n=3.
        \]
       Let us assume that $n=2$. By using that $u_2p_2\in \mathbb{Q}[\mathbf{y}]$ and $\gcd(u_2,\n)=1$, we obtain that $p_2\in \mathbb{Q}[\mathbf{y}]$.
         Similarly, since $q_2A_2\in \mathbb{Q}[\mathbf{y}]$ and $\gcd(A_2,\n)=1$ we deduce that $q_2\in \mathbb{Q}[\mathbf{y}]$.
        Same reasoning holds to show the existence of $p_3$ and $q_3$ in $\mathbb{Q}[\mathbf{y}]$ for the $n=3$ case.
\end{proof}
 
We define the sets
\begin{equation}
\begin{aligned}
\mathcal{W}_1^{(\epsilon)}&=\mathcal{W}_{A_1}\cap\{\epsilon B_1\leq 0\}\\
\mathcal{W}_k^{(\epsilon)}&=
\mathcal{W}_{A_k}\cap \{\epsilon p_k\geq 0\}),\quad\forall k\in J_n
\\
\mathcal{W}_\infty^{(\epsilon)}&=\cup_{k=1}^n\mathcal{W}_k^{(\epsilon)}\\
F_\epsilon&=
F_{o}\backslash \mathcal{W}_\infty^{(\epsilon)}. \\
\end{aligned}
\label{eq:defF_espetal}
\end{equation}
where we  $\mathcal{W}_{A_k}=\mathbf{V}_\mathbb{R}(A_k)\cap F_o$.
\noindent The following hypothesis leads to finite fibers.% on  the polynomials $(A_k)_{1\leq k\leq n},B_1,C_0,\n,(u_k)_{k\in J_n}, (p_{k})_{k\in J_n}$ of $\mathbb{Q}[\mathbf{y}]$:
\begin{hyp}
    \label{hyp:AkC0}
There exists a set $E_o$ such that $\mathbb{R}^{t+n}\supset E_o\supset E$ and the following holds:
\begin{equation}
(\mathcal{H})\left\lbrace
    \begin{aligned}
        &\textrm{(a)}\quad E_o\cap \mathbf{V}(C_0)=\emptyset\\
        &\textrm{(b)}\quad \forall k\in J_n\quad  E_o\cap \mathbf{V}(A_1,A_k)=\emptyset  \\
        &\textrm{(c)}\quad E_o\cap \mathbf{V}(\n)=\emptyset\\
        &\textrm{(d)}\quad  E_o\cap \mathbf{V}(A_1,B_1)=\emptyset\\
        &\textrm{(e)}\quad \forall k\in J_n\quad E_o\cap \mathbf{V}(A_k,p_k)=\emptyset\\
        &\textrm{(f)}\quad \forall k\in J_n\quad E_o\cap \mathbf{V}(A_k,u_k)=\emptyset \\
    \end{aligned}
    \right.
\end{equation}
\end{hyp}
This following theorem is the main result of this subsection.
\begin{theo}
\label{theo:xi1epsContinuousThroughA1epsB1}
    Grant \cref{hyp:codim} and \cref{hyp:AkC0}, $\xi^{(\epsilon)}:F_\epsilon\subset\mathbb{R}^t\to \xi^{(\epsilon)}(F_\epsilon)\subset\mathbb{R}^n$ is continuous, that is, for each $C\in \cc{F_\epsilon}$, $\xi^{(\epsilon)}$ is continuous from $C$ to $\xi^{(\epsilon)}(C)$
and the following holds:
\[
E=\cup_{\epsilon\in \{-1,1\}}E^{(\epsilon)}\hquad 
\text{ with }\hquad E^{(\epsilon)}=\{(\mathbf{y},\mathbf{x})\in E,\ \mathbf{y}\in F_\epsilon, \ \epsilon( 2A_1x_1+B_1)\geq 0\}\hquad
%\overset{def}{=} E\cap \Gamma(\xi^{(\epsilon)}_{|F_\epsilon})=\{(\mathbf{y},\mathbf{x})\in E,\ \mathbf{y}\in F_\epsilon, \ \epsilon( 2A_1x_1+B_1)\geq 0\}
\]
and $ E^{(\epsilon)}\cong \pi(E^{(\epsilon)})$ that is $(I\times\xi^{(\epsilon)})\circ \pi_{|E^{(\epsilon)}} =I_{|E^{(\epsilon)}}$.
\end{theo}
\begin{proof}
    See Appendix~\ref{sec:appendixProofContinuity}.
\end{proof}
\begin{coro}
    Grant \cref{hyp:codim} and  \cref{hyp:AkC0}, and assuming that $E$ is of the form $E = \{(\mathbf{y},\mathbf{x})\in \mathbf{V}_\mathbb{R}(\mathbf{f})\ : \ \mathcal{G}_{oo}(\mathbf{y})\}$ with $\mathcal{G}_{oo}=\land_1^rg_k\sigma_k0$ where $\sigma_k\in \{<,\neq\}$, then \cref{theo:xi1epsContinuousThroughA1epsB1} is equivalent to states that
    $E^{(\epsilon)}\cong F_\epsilon$ with $F_o$  given by $F_o = \set{\Delta\geq 0\land \mathcal{G}_{oo}}\subset\mathbb{R}^t$.
\end{coro}
\noindent  We denote by 
$
E_\mathcal{H} = \{(\mathbf{y},\mathbf{x})\in E,\ \mathbf{y}\in \mathcal{W}_\mathcal{H}\}$ and $\wt{\mathcal{V}}=\mathbf{V}(\ftilde)$.
%E^{(\epsilon)} &= E\cap \Gamma(\xi^{(\epsilon)}),\ \epsilon\in \{-1,1\}
% E_\epsilon &= E\cap\Gamma(\xi^{(\epsilon)})= \{(\mathbf{y},\mathbf{x})\in E,\ \mathbf{y}\in F_\epsilon,\ \epsilon(2A_1x_1+B_1)\geq 0\}\\
\begin{lemma}
    \label{le:SingularPOints}
    Grant \cref{hyp:codim} and \cref{hyp:AkC0}, the singular points of $\pi$ restricted to $E$  defined by $\mathcal{K}(\pi,E)\overset{\text{def}}{:=}
    \{\mathbf{z}\in E\ : \ d_\mathbf{z}\pi T_\mathbf{z}\wt{\mathcal{V}}\subsetneq \mathbb{R}^t\}$ are given by $E_\mathcal{H}$
        and $\pi(\mathcal{K}(\pi,E))\subset\mathcal{W}_\mathcal{H}$.
\end{lemma}
\begin{proof}
    By writing that
$        \mathcal{K}(\pi,E)=\{(\mathbf{y},\mathbf{x})\in E\ : \ \textrm{rank}(\textrm{jac}_\mathbf{x}(\ftilde))<n\}$
     and writing that all the $n$-th minors of  $\textrm{jac}_\mathbf{x}(\ftilde)$ vanish and using \cref{hyp:codim}-(c), we get 
        $\mathcal{K}(\pi,E) =\{(\mathbf{y},\mathbf{x})\in E\ : \ 2A_1x_1+B_1=0\} = \{(\mathbf{y},\mathbf{x})\in E\ : \ \Delta=0\}=E_\mathcal{H}$.
\end{proof}
The immediate consequence of \cref{theo:xi1epsContinuousThroughA1epsB1} and \cref{le:SingularPOints}  is 
\begin{coro}
    \label{coro:ccIntersectInWh}
    Grant \cref{hyp:codim} and \cref{hyp:AkC0},
    each connected components of $E$ write as a finite union of those of $E^{(\epsilon)}$ for $\epsilon\in\{-1,1\}$ which can intersect on $E_\mathcal{H}$.
\end{coro}
\begin{proof}
% Let be $\epsilon\in \{-1,1\}$ and $C_k^{(\epsilon)}\in \cc{\Gamma(\xi^{(\epsilon)})}\cap E$, the fact that $\xi^{(\epsilon)}$ is continuous from $F_\epsilon$ into $\Gamma(\xi^{(\epsilon)})$ implies that there exists $C\in \cc{E}$ such that $C_k^{(\epsilon)}\subset C$.
Let  $C\in \cc{E}$, the number of connected components of a semi-algebraic set being finite  (\cite{basubook}-Theorem~5.21) we can define
 \[
\widehat{C} = \bigcup_{\epsilon\in \{-1,1\}}\bigcup_{ C'\in \ccpetit{E^{(\epsilon)}}\atop C'\subset C} C'
\]
Clearly $\widehat{C}\subset C$; let us assume that there exists $(\mathbf{y},\mathbf{x})\in {C}\backslash \widehat{C}$, then by \cref{theo:xi1epsContinuousThroughA1epsB1}, there exists $\epsilon\in \{-1,1\}$ such that $\mathbf{x}=\xi^{(\epsilon)}(\mathbf{y})$ and $\widetilde{C}\in \cc{E^{(\epsilon)}}$ such that  $(\mathbf{y},\xi^{(\epsilon)}(\mathbf{y}))\in \widetilde{C}$.
 We deduce that $\widetilde{C}\not\subset C$ otherwise it would belong to $\widehat{C}$, hence the contradiction with the maximality of $C$.
The fact that  the pairwise intersection between connected components of the the distinct branches is included in $E_\mathcal{H}$ comes from the fact that 
\begin{align*}
    (\eta,\chi)\in E&\Longleftrightarrow\exists\epsilon\in \{-1,1\}\ : \ \chi =\xi^{(\epsilon)}(\eta)\\
    (\eta,\chi)\in E\land (\chi=\xi^{(-1)}(\eta)=\xi^{(1)}(\eta))
    &\overset{\textrm{\cite{basubook} Prop 4.96}}{\Longleftrightarrow}\chi \textrm{ is a singular point of } \ftilde(\eta,.)\\
    &\overset{\textrm{Lemma~\ref{le:SingularPOints}}}{\Longleftrightarrow}(\eta,\chi)\in E_\mathcal{H}\\
    \end{align*}
\end{proof}
%\noindent We deduce that the connected components of $E$ can be obtained from the ones of  $E^{(\epsilon)}$ for $\epsilon\in \{-1,1\}$ via a recursive merging  through $E_\mathcal{H}$.
%The connected components of $E^{(\epsilon)}$ are candidate for the connected components of $E$ though a recursive merging must be done through $E_\mathcal{H}$. \\
Let us denote by $I = \langle \ftilde\rangle$ and $\mathcal{V}_{Q_k}=\mathbf{V}(I+\langle Q_k\rangle)$.
%$\overline{Q_k}$ the representant of $Q_k$ in $\mathbb{Q}[\mathbf{y},\mathbf{x}]/I$.
\begin{hyp}
    \label{hyp:NoralFormQk}
    There exists $E_o\supset E$ s.t for all $k\in \mathcal{J}$, $\mathbf{V}(I+\langle Q_k\rangle)\cap E_o=\mathbf{V}(q_k,\alpha_kx_1+\beta_k,\tilde{f}_2,\dots \tilde{f}_{n+1})\cap E_o$ with $q_k\in \mathbb{Q}[\mathbf{y}]$ and $(\alpha_k,\beta_k)\in \mathbb{Q}[\mathbf{y}]^2$ and $E_o\cap \mathbf{V}(\alpha_k)=\emptyset$.
\end{hyp}
%Let $k\in \mathcal{J}$, to shorten notations we assume that $\overline{Q_k}=\alpha_kx_1+\beta_k$.
\begin{lemma}\label{le:expressionVQk}
    Grant \cref{hyp:codim}, \cref{hyp:AkC0}, \cref{hyp:NoralFormQk}, and assume that $E_o$ is of the form, $E_o = \{(\mathbf{y},\mathbf{x})\in \mathbb{R}^{t+n}\ :\ \mathcal{G}_{oo}(\mathbf{y})\}$, 
    let   $\mathcal{V}_{Q_k}^{(\epsilon)}=\mathcal{V}_{Q_k}\cap \Gamma(\xi^{(\epsilon)}) $, with $\xi^{(\epsilon)}:F_\epsilon\to \mathbb{R}^{t+n}$ with $F_\epsilon$ defined by \eqref{eq:defF_espetal} by setting $F_o=\set{\Delta\geq 0\land \mathcal{G}_{oo}}$, then for both the cases $n\in\{2,3\}$:
\begin{equation}
    \label{eq:expressionVQk}
     \mathcal{V}_{Q_k}^{(\epsilon)}=\{(\mathbf{y},\xi^{(\epsilon)}(\mathbf{y})), \ \mathbf{y}\in F_\epsilon\ : \    (q_k(\mathbf{y}) =0)\land (\epsilon\underbrace{\alpha_k(-2A_1\beta_k+B_1\alpha_k)}_{\overset{\text{def}}{=}-s_k(\mathbf{y})}\geq 0)\}
\end{equation} 
%where $F_\epsilon$ is given by \eqref{eq:defF_espetal} with $F_o=\set{\Delta\geq0\land \mathcal{G}_o}$. Besides $\mathcal{V}_{Q_k}^{(\epsilon)}$
and projects as 
\begin{equation}
    \label{eq:projectionVQkeps}
\pi(\mathcal{V}_{Q_k}^{(\epsilon)})=\{\mathbf{y}\in F_\epsilon\ :(q_k(\mathbf{y})=0)\land(\epsilon s_k(\mathbf{y})\leq0)\}.
\end{equation}
\end{lemma}
\begin{proof}
Let $H_k^{(\epsilon)}$ be the set of the right-hand side of \eqref{eq:expressionVQk}.  
\begin{align*}
        (\mathbf{y},\mathbf{x})\in \mathcal{V}_{Q_k}^{(\epsilon)}&\Longleftrightarrow \mathbf{y}\in F_\epsilon\land (\mathbf{x}=\xi^{(\epsilon)}(\mathbf{y}))\land  (Q_k(\mathbf{y},\mathbf{x})=0)\\
        &\overset{\cref{theo:xi1epsContinuousThroughA1epsB1}}{\Longleftrightarrow} \mathbf{y}\in F_\epsilon\land (\tilde{f}(\mathbf{y},\mathbf{x})=0)\land (\epsilon(2Ax_1+B_1) \geq 0) \land (Q_k(\mathbf{y},\mathbf{x})=0)\\
        &\overset{\cref{hyp:NoralFormQk}}{\Longleftrightarrow}\mathbf{y}\in F_\epsilon\land \alpha_kx_1+\beta_1=\wt{f}_2(\mathbf{y},\mathbf{x})=\dots=\wt{f}_{n+1}(\mathbf{y},\mathbf{x})=0\land q_k(\mathbf{y})=0\\
        &\hspace{3cm}\land (\epsilon(2Ax_1+B_1) \geq 0) \\
        & \Longleftrightarrow(\mathbf{y},\mathbf{x})\in H_k^{(\epsilon)}
    \end{align*}
    % Let $(\mathbf{y},\mathbf{x})\in H_k^{(\epsilon)}$, from Assumption~\ref{hyp:NoralFormQk}, we deduce that $\tilde{f}(\mathbf{y},\mathbf{x})=0$ and $Q_k(\mathbf{y},\mathbf{x})=0$.
    % From Theorem~\ref{theo:xi1epsContinuousThroughA1epsB1} we deduce that   $\mathbf{x}=\xi^{(\epsilon)}(\mathbf{y})$ and $(\mathbf{y},\mathbf{x})\in E^{(\epsilon)}$ hence verifies the inequality $\epsilon(2Ax_1+B_1) \geq 0$ that is equivalent over $E\subset E_o$ to  $\epsilon\alpha_k(-2A_1\beta_k+B_1\alpha_k)\geq0)$ hence the inclusion $H_k^{(\epsilon)}\subset \mathcal{V}_{Q_k}^{(\epsilon)}$.
    % Conversely, let $(\mathbf{y},\mathbf{x})\in \mathcal{V}_{Q_k}^{(\epsilon)}$ then $\mathbf{x}=\xi^{(\epsilon)}(\mathbf{y})$ verifies $\tilde{f}(\mathbf{y},\mathbf{x})=0$ and $Q_k(\mathbf{y},\mathbf{x})=0$. By \cref{hyp:NoralFormQk}, $q_k(\mathbf{y})=0$ and $x_1 = -\frac{\beta_1}{\alpha_1}$. Besides, from \cref{theo:xi1epsContinuousThroughA1epsB1}, $(\mathbf{y},\mathbf{x})\in E^{(\epsilon)}$ and verifies  $\epsilon(2Ax_1+B_1) \geq 0$ which rewri

    % Hence $\tilde{f}(\mathbf{y},\mathbf{x})=0$ and as $q_k(\mathbf{y})=0$ thanks to \cref{hyp:NoralFormQk}, 
\end{proof}
In the sequel, we assume that there exists $E_o$ satisfying \cref{hyp:codim}, \cref{hyp:AkC0} and \cref{hyp:NoralFormQk} and of the form given in \cref{le:expressionVQk} with $\mathcal{G}_{oo}:\mathbb{R}^t\to\{0,1\}$.
We abuse notation and denote by $F_\epsilon$ the set  defined in \eqref{eq:defF_espetal} by replacing $F_o$ by  $F_o=\set{\Delta\geq0\land \mathcal{G}_{oo}}$.
We denote by $\mathcal{W}_{Q_k}^{(\epsilon)}=\pi(\mathcal{V}_{Q_k}^{(\epsilon)})$, $G_\epsilon=F_\epsilon\backslash \cup_{1\leq k\leq N} \mathcal{W}_{Q_k}^{(\epsilon)}$ and
\begin{align}
    L&\leftarrow\texttt{Sampling}(\{(\Delta>0)\land(w_\infty w_Q\neq 0)\}),\nonumber\\
    L_o^{(\epsilon)}&=\{\eta\in L, \ (\eta,\xi^{(\epsilon)}(\eta))\in E\},\nonumber\\
    G_{o,\epsilon}&=\bigcup_{C\in \ccpetit{G_\epsilon}\atop  C\cap L_o^{(\epsilon)}\neq \emptyset} C.
    \label{eq:defGoeps}
    \end{align}
 where \texttt{Sampling} denotes an exhaustive semi-algebraic set sampling algorithm (\texttt{CAD} (see \cite{Collins:75}) or more recent algorithms based on Morse Theory (see \cite{schostElDin2003} and \cite{LE202225}-Corollary~3)).
\begin{theo}

    \label{theo:CCdeGepsetCCdeE}
    Grant \cref{hyp:codim}, \cref{hyp:AkC0} and \cref{hyp:NoralFormQk}, if for all $k\in \{1,...r\}$ $\sigma_k\in \{\neq,>\}$, then $E^{(\epsilon)}$ is homeomorphic to $G_{o,\epsilon}$.
    % each $\widehat{C}\in \cc{E\cap \Gamma(\xi^{(\epsilon)})}$ is homeomorphic to a $C\in \cc{G_\epsilon}$.
\end{theo}
\begin{proof}
    Let us start by remarking that since $\xi^{(\epsilon)}$ is continuous on $F_\epsilon$ (see \cref{theo:xi1epsContinuousThroughA1epsB1}), then $\Gamma(\xi^{(\epsilon)}_{|F_\epsilon})$  is homeomorphic to $F_\epsilon$ via $h_\epsilon(\mathbf{y})=(\mathbf{y},\xi^{(\epsilon)}(\mathbf{y}))$ bi-continuous from $F_\epsilon$ into $\Gamma(\xi^{(\epsilon)}_{|F_\epsilon})$.\\
%    Let us note that this fact is in correspondance with the Hardt's triviality theorem (Theorem~\ref{theo:extensionTheorem}).
%     \\    
    %where  $Sampling(X)$ is an algorithm which gives at least one point point by connected component of  the  semi-algebraic set $X$ (see \cite{wolfram2023semi-algebraiccomponentinstances} and \cite{LE202225}-Corollary~3 for  theoretical results about existence and complexity of such algorithm).
    Since $G_{o,\epsilon}\subset F_\epsilon$, $\xi^{(\epsilon)}:G_{o,\epsilon}\to \xi^{(\epsilon)}(G_{o,\epsilon})$ is continuous and  for each $C\in \cc{G_{o,\epsilon}}$, $\xi^{(\epsilon)}(C)$ does not meet $\mathcal{V}_{Q}^{(\epsilon)}=\cup_{k\in \mathcal{J}}\mathcal{V}_{Q_k}^{(\epsilon)}$. 
    Hence $\xi^{(\epsilon)}(C)$ is connected and as there exists $\eta\in L_{o}^{(\epsilon)}\cap C$ we deduce that $\Gamma(\xi^{(\epsilon)}_{|C})\subset E$ and is homeomorphic to $C$ via $h_\epsilon$.
    Hence by finite union, $\Gamma(\xi^{(\epsilon)}_{|G_{o,\epsilon}})\subset E\cap\Gamma(\xi^{(\epsilon)}_{|F_\epsilon})=E^{(\epsilon)}$.
    Let us now prove that $E\cap\Gamma(\xi^{(\epsilon)}_{|F_\epsilon})\subset \Gamma(\xi^{(\epsilon)}_{|G_{o,\epsilon}})$. 
    Let us assume that there is $\eta\in F_\epsilon$ such that $({\eta},\xi^{(\epsilon)}(\eta))\in (E\cap\Gamma(\xi^{(\epsilon)}_{|F_\epsilon}))\backslash \Gamma(\xi^{(\epsilon)}_{|G_{o,\epsilon}}) $, then $({\eta},\xi^{(\epsilon)}(\eta))\not\in \mathcal{V}_{Q}^{(\epsilon)}$ and there exists $C\in \cc{G_{\epsilon}}$ such that $C\ni \eta$.
    Since $\mathcal{W}_\infty$ and $\mathcal{W}_Q$ are sets of empty interior in $\mathbb{R}^t$ and $\xi^{(\epsilon)}$ is continuous around $\eta$, the set $C$ contains a connected component of $\{\Delta>0,\ (w_\infty w_Q)\neq 0\}$.
    As \texttt{Sampling} is exhaustive,  $C\subset G_{o,\epsilon}$  hence the contradiction. 
    Finally, we have shown that $\Gamma(\xi^{(\epsilon)}_{|G_{o,\epsilon}})=E\cap\Gamma(\xi^{(\epsilon)}_{|F_\epsilon})$ and that each $C\in \cc{G_{o,\epsilon}}$ is homeomorphic to $\Gamma(\xi^{(\epsilon)}_{|C})$. 
    The fact that $\Gamma(\xi^{(\epsilon)}_{|C})$ is maximal for the inclusion and belongs to $ \cc{\Gamma(\xi^{(\epsilon)}_{|G_{o,\epsilon}})}$ comes from the fact that  $\pi(E^{(\epsilon)})\subset F_\epsilon$   and that $\pi(\mathcal{V}_{Q_k}^{(\epsilon)})=\mathcal{W}_{Q_k}^{(\epsilon)}$ for all $k\in \mathcal{J}$.
    % \[
    % (\mathbf{y},\mathbf{x})\in E\cap\Gamma(\xi^{(\epsilon)})\Longleftrightarrow  \mathbf{y}\in G_{o,\epsilon}\land \mathbf{x}=\xi^{(\epsilon)}(\mathbf{y})
    % \]

\end{proof}

\paragraph{Algorithm to compute $G_{o,\epsilon}$}
Let us assume that $\mathcal{G}(\mathbf{y},\mathbf{x})=\mathcal{G}_{oo}(\mathbf{y})\land \widetilde{\mathcal{G}}(\mathbf{y},\mathbf{x})$.
 Up to replace $F_o$ by $ \set{\Delta\geq 0\land \mathcal{G}_{oo}}$, we assume that  $\mathbf{g}=(g_1,...,g_r)\in \mathbb{Q}[\mathbf{y},\mathbf{x}]^r$ and $\bm{\sigma} \in \{<,\neq\}^r$ define the logical formula  $\widetilde{\mathcal{G}}=\mathbf{g}\sigma 0$.
 Let $L_X$ be a list of semi-algebraic connected sub-graphs of $\mathbb{R}^t\times \mathbb{R}^n$ and  $W\subset\mathbb{R}^t$ a semi-algebraic set, we denote by $\texttt{Merge}(L_X,W)$  an algorithm which merges recursively a pair of elements of $L_X$ whose the projection intersects in $W$.
  Namely, for $(X_i,X_j)\in L_X^2$ if $\pi(X_{i})\cap \pi(X_{j})\cap W\neq \emptyset$ then $X_{i}$ and $X_{j}$ are merged in one set $X_i\cup X_j$. This operation is repeated while no pairs in $L_X$ intersects in $W$. 
  Such algorithm exists in particular if the intersection between the frontiers of the elements of $L_X$  and $W$ are known in closed form.
  Grant \cref{hyp:codim}, \cref{hyp:AkC0} and \cref{hyp:NoralFormQk},
  \cref{theo:xi1epsContinuousThroughA1epsB1}, \cref{theo:CCdeGepsetCCdeE} and \cref{coro:ccIntersectInWh} hold and we deduce that \cref{alg:computeGeps} enables to describe  the connected components of $E$ from the ones of $\pi(E^{(\epsilon)})$ for $\epsilon\in \{-1,1\}$. 
\begin{algorithm}[H]

\caption{\label{alg:computeGeps}Computation of $(G_{o,\epsilon)_{\epsilon=\pm 1}}$ and $\cc{E}$}
%  \begin{algorithmic}
%     \item 
 \begin{align*}
 F_o&=\{\mathbf{y}\in \mathbb{R}^t\ : \ \mathcal{G}_{oo}(\mathbf{y})\land \Delta(\mathbf{y})\geq 0 \} \text{ and }\mathcal{W}_k^{(\epsilon)}, \ \mathcal{W}_\infty^{(\epsilon)},\ F_\epsilon \text{ given by } \eqref{eq:defF_espetal}\\
% \mathcal{W}_1^{(\epsilon)}&=\mathcal{W}_{A_1}\cap\{\epsilon B_1\leq 0\}\\
% \mathcal{W}_k^{(\epsilon)}&=
% \mathcal{W}_{A_k}\cap \{\epsilon p_k\geq 0\}\quad \forall k\in J_n
% \\
% \mathcal{W}_\infty^{(\epsilon)}&=\cup_{k=1}^n\mathcal{W}_k^{(\epsilon)}\\
% F_\epsilon&=
% F_{o}\backslash \mathcal{W}_\infty^{(\epsilon)} \\
%\mathcal{W}_{Q_l}^{(\epsilon)}&=\{\mathbf{y}\in F_\epsilon\ : \ (q_l(\mathbf{y})=0)\land(\epsilon\alpha_l(-2A_1\beta_l+B_1\alpha_l)\geq0)\} \quad \text{ for }l\in \mathcal{J}\\
\mathcal{W}_{Q_l}^{(\epsilon)}&\text{ given by }\eqref{eq:projectionVQkeps}
\text{ and }G_\epsilon=F_\epsilon\backslash\cup_{l\in \mathcal{J}} \mathcal{W}_{Q_l}^{(\epsilon)}\\
L&\leftarrow\texttt{Sampling}(\{(\Delta>0)\land(w_\infty w_Q\neq 0)\})\text{ and }
    L_o^{(\epsilon)}=\{\eta\in L, \ (\eta,\xi^{(\epsilon)}(\eta))\in E\}\\
    G_{o,\epsilon}&=\bigcup_{C\in \ccpetit{G_\epsilon}\atop  C\cap L_o^{(\epsilon)}\neq\emptyset} C\text{ and }
    \CC^{(\epsilon)}=\{\Gamma(\xi ^{(\epsilon)}_{|C}),\ C\in \ccpetit{G_{o,\epsilon}}\}\\
    \pi_0(E)&\leftarrow\texttt{Merge}(\cup_{\epsilon\in \{-1,1\}}\CC^{(\epsilon)},\mathcal{W}_\mathcal{H})
\end{align*}
% \end{algorithmic}
\end{algorithm}

% \begin{re}
%     \label{re:algorithmToComputeCC}
%     If the elements of $(\CC^{(\epsilon)})_{\epsilon=\pm 1}$ are not known in closed form, the last step of Algorithm~\ref{alg:computeGeps} consisting in merging the connected set of $\CC^{(-1)}$ and $\CC^{(1)}$  whose their respective projection have a common intersection  lying in $\mathcal{W}_\mathcal{H}$ is compromised.
%     Nevertheless, Algorithm~\ref{alg:computeGeps} gives a characterisation of the connected components of $E$ in function of the connected components of a  lower dimensional semi-algebraic set  involving only  inequalities.
% \end{re}

\paragraph{Topological invariant} To identify the connected components we introduce  the notion of topological invariant as
%Let us introduce the notion of topological invariant   useful to characterize the connected components of a semi-algebraic set.
\begin{defi}
   \label{de:topologicalInvariant}
   Let $X$ be a topological space and $N$ be a countable set, we say that $\mathcal{S}:X\to N$  is a topological invariant over $X$ if 
   \[ \forall (x,x')\in X^2\ [\exists \gamma \in C^0([0,1],X)\ : \  \gamma(0)=x\land \gamma(1)=x']\Longrightarrow \mathcal{S}(x)=\mathcal{S}(x')
       \]
   %In other word $\mathcal{S}$ is a topological invariant if it factorizes as $\mathcal{S}=i\circ \mathcal{N}$ with $i:\cc{X} \to N$ a function and $\mathcal{N}:X\to \cc{X}$  the representant map for the homotopy equivalence $\overset{X}{\sim}$ (see \eqref{eq:homotopyEquivalence}).
   We say that $\mathcal{S}$ is exact if we replace the implication by an equivalence.
\end{defi}
% \begin{re}
% \label{re:topologicalInvariantDef}
% Two connected components of $E$ can have the same topological invariant, especially if this invariant is not exact.
% However, two connected sets corresponding to different values of a topological invariant cannot belong to the same connected component.
% \end{re}
\begin{lemma}
\label{le:topologicalInvariantCasGeneral}
Let us assume that $\sigma_k\in \{>,\neq\}$ for all $k\in \llbracket 1,r\rrbracket$ and
let  $r'\leq r$  such that $ \{i_1,...,i_{r'}\}\subset\llbracket 1,\rrbracket$ such that $\sigma_{k'}\in \{\neq\}$ for all $k'\in \{i_1,...,i_{r'}\}$.
Then, $\mathcal{S} :E\to \{-1,1\}^{2+r'}$ defined by $ \mathcal{S}(\mathbf{y},\mathbf{x})=\sign{[x_{n-1},x_n,g_{i_1}(\mathbf{y},\mathbf{x}),...,g_{i_{r'}}(\mathbf{y},\mathbf{x})]}$ is a topological invariant over $E$.
\end{lemma}
\begin{proof}
Grant $\mathcal{H}-(a)$ of \cref{hyp:AkC0}, the product $x_{n-1}x_n$ does not cancel out over $E$, hence $\mathcal{S}_1=\sign{[x_{n-1},x_n]}$ is a topological invariant over  $E$.
 Similarly, since $\sigma_k\in \{\neq,<\}$ the product   $\prod_{k=1}^rg_k$ does not vanish over $E$. By keeping only the $g_k$ associated to the operator $\neq$, we define the topological invariant $\mathcal{S}_2=\sign{[g_{i_1},...,g_{i_{r'}}]}$ and we deduce that $\mathcal{S}=\mathcal{S}_1\times \mathcal{S}_2$ is a topological invariant over $E$.
%   out, making $\wt{\mathcal{S}_2}=\sign{[g_1,...,g_r]}$  a topological invariant over  $E$.
%    Since for all $k\in \{1,...r\}\backslash\{i_1,...,i_{r'}\}$, the sign of $g_k$ remains constant over $E$ , these terms become irrelevant. Thus, we define the topological invariant  
\end{proof}

  \subsection{Application to optics}
  \label{sec:optical_application}
  Let us start by defining the  \textit{optical admissible solutions} set  $E$. 
Let $N\geq 3$ be the number of mirrors, $n\in \{2,3\}$ the number of polynomial equations.
We still denote by $\mathbf{y}\in\mathbb{R}^t$  and $\mathbf{x}\in\mathbb{R}^n$ such that the  union of  $\mathbf{y}$ and $\mathbf{x}$ corresponds to the unknowns described in \cref{sec:polynomialFocalSystem} and \cref{sec:polynomialAFocalSystem}.
Let $\pi:\mathbb{R}^{t+n}\ni(\mathbf{y},\mathbf{x}) \mapsto\mathbf{y}\in \mathbb{R}^t$ the canonical projection,
and $\f\in\mathbb{Q}[\mathbf{y},\mathbf{x}]^n$ the polynomial sequence corresponding to either $\mathbf{g}_{n,N}$ (see \cref{sec:polynomialFocalSystem}) or $\mathbf{h}_{n,N}$ (see \cref{sec:polynomialAFocalSystem}),
$n$ will be frequently named as the codimension of the associated  algebraic set $\mathbf{V}(\f)\subset \mathbb{C}^{t+n}$. 
 Let us recall that $\Go(\mathbf{y},\mathbf{x})$  is the  set of constraints  given by \eqref{eq:constraintFocal} and \eqref{eq:constraintAFocal} for respectively focal and afocal telescopes.
 We assume that \(\Go(\mathbf{y}, \mathbf{x})\) can be decomposed as
\[
\Go(\mathbf{y}, \mathbf{x}) = \Goo(\mathbf{y}) \land \Ho (\mathbf{y}, \mathbf{x}),
\]
and we define  $\Goohat$ the logical clause induced by $\Goo$ over $\mathbb{R}^{t+n}$ such that  \(\Goohat(\mathbf{y}, \cdot) = \Goo(\mathbf{y})\).
We set   $E_o =  V_\mathbb{R}(\f)\cap \set{\Go}$, $E_{oo} =  \set{\Goohat}$, and $F_{oo} =  \set{\Goo} $.
Each symbolic quantity $\Phi$ depending explicitly on $(\mathbf{y},\mathbf{x})$ are  abusely denoted as $\Phi(\mathbf{y},\mathbf{x})$.
 For $k\in \llbracket 1,N\rrbracket$, let  $Q_k\in \mathbb{Q}[\mathbf{y},\mathbf{x}]$ be the polynomial such that for all $ (\mathbf{y},\mathbf{x})\in E_{o}$:
 \begin{equation}
\begin{aligned}
\label{eq:equivalenceCOurbureQk}c_k(\mathbf{y},\mathbf{x})=0 &\Longleftrightarrow Q_k(\mathbf{y},\mathbf{x})=0.
\end{aligned}\end{equation}
%, $\mathcal{S}_2=\sign([c_1,...,c_k])$ for $k\in \{1,...,N\}
These polynomials are given by\footnote{The two conditions which are used here are $f_1=0$ and $\mathcal{G}_o$.} 
\begin{align*}
    Q_1&=\Omega_1-1,\hquad 
    Q_k=\Omega_{k-1}(d_{k-1}-d_k)+d_k-\Omega_k\Omega_{k-1}d_{k-1}\hquad \text{for}\hquad k\in \llbracket 2, N-1\rrbracket\\
    Q_N&=\left\lbrace \begin{array}{cc}
        (1-\Omega_{N-1})d_N+\Omega_{N-1}d_{N-1},& \text{Focal case}\\
        \Omega_{N-1}-1,& \text{ Afocal case}
    \end{array}\right.&
\end{align*}
and we define 
\[
    \mathcal{G}=\Go\land \GQ\quad \text{ with }\quad \GQ=\bigwedge_{1\leq k\leq  N}(Q_k\neq 0).
    \]
\begin{defi}[Optical admissible solutions]
\label{def:admissbleSolution}
 
    We define the set of  optical admissible solutions as $E=\mathbf{V}_\mathbb{R}(\f)\cap \set{\G}= E_o\backslash\cup_{k=1}^N E_{Q_k}$,
        where $E_{Q_k}=\mathbf{V}_\mathbb{R}(\langle\f\rangle +\langle Q_k\rangle )\cap \set{\mathcal{G}_o}$.
        The set of optically admissible solutions consists of real, non-optically-degenerate real solutions to \(\f = 0\).
    \end{defi}

%  The sequence of the polynomial sequences for the focal  and afocal telescopes are given by $\mathbf{g}_{n,N}$ \eqref{eq:defpolynomialsequenceg}-\eqref{eq:summarizedEqFocal} and $\mathbf{h}_{n,N}$ \eqref{eq:defpolynomialsequenceh}-\eqref{eq:summarizedEqAfocal}, for $2\leq n\leq 3$.
%   In the sequel of this section, we assume that $E$ is given by one of these polynomial sequences / constraints defining the focal and afocal telescopes.
To name the connected components of \( E \), we introduce a nomenclature based on the signature of the vector formed by magnifications and curvatures, which serves as a topological invariant over \( E \) (see \cref{de:topologicalInvariant} and \cref{le:nomenclature}):
 \begin{defi}
     \label{def:nomenclature}
     Let  $\mathbf{c}:\mathbb{R}^{t+n}\ni(\mathbf{y},\mathbf{x})\mapsto (c_1,...,c_N)\in\mathbb{R}^N$ (see \eqref{eq:courbures}),  $\bm{\Omega}:\mathbb{R}^{t+n}\ni(\mathbf{y},\mathbf{x})\mapsto (\Omega_1,...,\Omega_{N-1})\in\mathbb{R}^{N-1}$ and
     \[
         \Psi:
         \left\lbrace
         \begin{aligned}
            \mathbb{R}^{N-1}\times \mathbb{R}^{N}&\longrightarrow \{-1,1\}^{2N-1}\\
             (\mathbf{a},\mathbf{b})&\mapsto ((\sign(a_k))_k,((-1)^k\sign(b_k))_k).
         \end{aligned}
         \right.
         \]  
         We introduce $\mathcal{S}:E\to\{0,1\}^{2N-1}$ the signature of the set of magnifications and curvatures:
         \[
            \mathcal{S}(\mathbf{y},\mathbf{x})=\Psi(\bm{\Omega}(\mathbf{y},\mathbf{x}),\mathbf{c}(\mathbf{y},\mathbf{x})).
             \]
         In order to distinguish the magnifications and the curvatures sign in the name, we put the letter $\texttt{P}$ (resp. digit $\texttt{1}$) when magnification (resp. curvature) is positive and $\texttt{N}$ (resp. digit $\texttt{0}$) otherwise.
        
 \end{defi}

\begin{figure}[htbp]
    \begin{center}
        \includegraphics[width=0.7\textwidth]{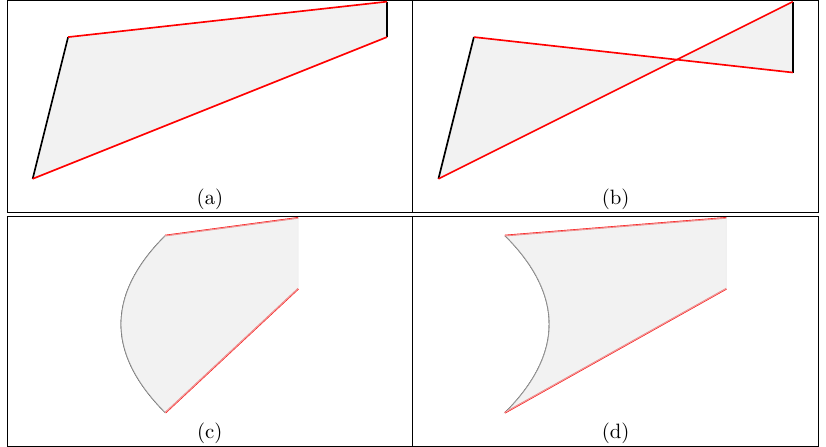}
\caption{\label{tab:topo_invariant}Topological invariant features. (a) Positive and (b) Negative magnifications. (c) Convex and (d) Concave mirrors.}
\end{center}
\end{figure}

 \begin{re}
     \label{re:nomenclature}
     For example, a typical nomenclature is $\texttt{PP101}$ what means that the two magnifications are positive and  $-c_1>0$ (convex), $c_2<0$ (concave), $-c_3>0$ (convex).
   \end{re}
   \begin{lemma}
    \label{le:nomenclature}
       $\mathcal{S}$ is a topological invariant over the set $E$ of \cref{def:admissbleSolution}.
   \end{lemma}
\begin{proof}
    Following the proof of   \cref{le:topologicalInvariantCasGeneral} and remarking that the product $\Omega_s$ does not vanish over $E$, and using \eqref{eq:equivalenceCOurbureQk} we get the result.
\end{proof}
The collection of  topological invariants that make up $\mathcal{S}$  are illustrated in \cref{tab:topo_invariant}.

\begin{re}
    \label{re:assumptionQk}
%\paragraph{About Assumption~\ref{hyp:NoralFormQk} for $\bm{N=3}$.}
 Grant \cref{hyp:codim}, \cref{hyp:NoralFormQk} is verified if there exists $(\mathcal{Q}_k)_{k\in \mathcal{J}}$ such that  
    \begin{equation}
    \label{eq:AssumptionV_QVerified}
    \forall k\in \mathcal{J}\ \ \mathbf{V}(\langle \ftilde\rangle+\langle {Q}_k\rangle)\cap E_{oo}=\mathbf{V}(\langle \ftilde\rangle+\langle \mathcal{Q}_k\rangle)\cap E_{oo}=\mathbf{V}(q_k,\alpha_k x_1+\beta_k,\tilde{f}_2,\dots,\tilde{f}_{n+1})\cap E_{oo}
    \end{equation}
    where $q_k\in \mathbb{Q}[\mathbf{y}] $ and $\alpha_k,\beta_k\in \mathbb{Q}[\mathbf{y}]$ are obtained by using  \cref{theo:eliminationTheorem} and a Groebner basis computation.
    For $N=3$, let us note that if the focal relation \(0 = \Omega_1 \Omega_2 f + d_3\) (see \eqref{eq:C1focale})  holds true, then \((Q_k)_{k\in\llbracket 1,3\rrbracket}\) can be substituted with \((\mathcal{Q}_k)_{k\in\llbracket 1,3\rrbracket}\) as follows:
    \begin{align}
        \mathcal{Q}_1= \Omega_1-1\quad \mathcal{Q}_2 = \Omega_1f(d_1-d_2)+d_2f+d_1d_3\quad
        \mathcal{Q}_3=\Omega_1f +d_3-d_2.\label{eq:Q_kFocal}
    \end{align}
    Analogously, if the magnification relation \(0 = \Omega_1 \Omega_2 - G\) (see \eqref{eq:C1grandissement}) holds true, then \(Q_k\) can be substituted with \(\mathcal{Q}_k\) as follows:
\begin{align}
    \mathcal{Q}_1=\Omega_1-1\,\quad\mathcal{Q}_2= \Omega_1(d_1-d_2)+d_2-Gd_1,\quad
    \mathcal{Q}_3=G-\Omega_1.\label{eq:Q_kAFocal}
\end{align}
\end{re}
% For codimension $2$, considering that $x_1$ is equal to $\Omega_1$, by identifying we easily get the expressions of the polynomials $\alpha_k$ and $\beta_k$ and the $q_k$ are obtained by replacing $x_1$ by $-\frac{\beta_k}{\alpha_k}$ in $f_1$ showing that Assumption~\ref{hyp:NoralFormQk} is verified.
% For codimension $3$, we will use 
\noindent In the sequel of the paper, we assume that $N=3$ and we use notations of \cref{sec:triangular_system}. 
We will verify \cref{hyp:codim}, \cref{hyp:AkC0} and \cref{hyp:NoralFormQk} in order to apply \cref{alg:computeGeps}.
When all the topological invariants associated with each connected set that composes $\CC^{(\epsilon)}$ are different, we denote it by the topological invariant with the branch indicated as a superscript.
% Let  $k\in \{1,\dots,n\}$, $j\in \{1,\dots,N\}$ and $\epsilon\in \{-1,1\}$, we recall that  $\mathcal{W}_{A_k}=\mathbf{V}_\mathbb{R}(A_k)\cap F_o$  with $F_o=\set{\Delta \geq 0\land \mathcal{G}_{oo} }$,
%  $\mathcal{V}_{Q_j}=\mathbf{V}_\mathbb{R}(\langle \ftilde\rangle+\langle \mathcal{Q}_j\rangle)$, $\mathcal{V}_{Q_j}^{(\epsilon)}=\mathcal{V}_{Q_j}\cap \set{\epsilon(2A_1x_1+B_1)\geq 0}$, $\mathcal{W}_{Q_j}^{(\epsilon)}=\pi(\mathcal{V}_{Q_j}^{(\epsilon)})$ and   $F_\epsilon$, $\mathcal{W}_\infty^{(\epsilon)}$ and $\mathcal{W}_k^{(\epsilon)}$ are given in \eqref{eq:defF_espetal}.

  \subsubsection{Systems of codimension $n=2$}
     
\begin{figure}[htbp]
    \begin{center}
  \includegraphics[width=\textwidth]{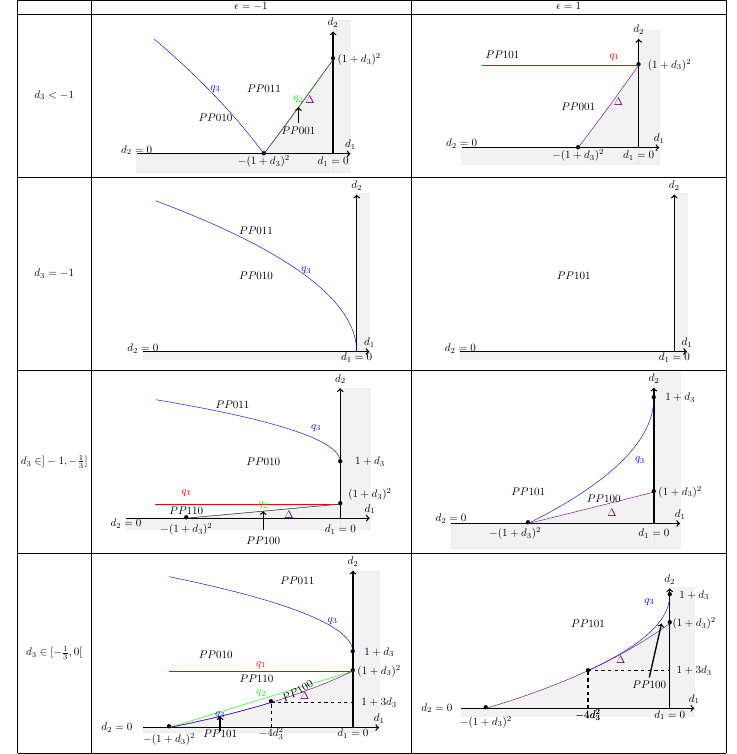}
  \caption{\label{tab:focal_codim2_fpos}Focal codim 2, $f=1$}
  \end{center}
  \end{figure}

\begin{figure}[htbp]
    \begin{center}
  \includegraphics[width=\textwidth]{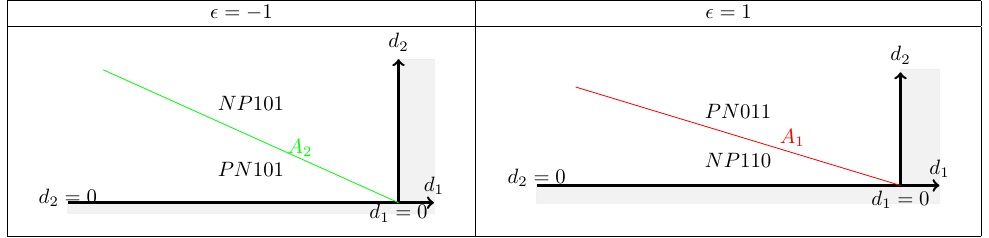}
  \caption{\label{tab:focal_codim2_fneg}Focal codim 2, $f=-1$}
  \end{center}  
  \end{figure}

\begin{figure}[htbp]
    \begin{center}
  \includegraphics[width=\textwidth]{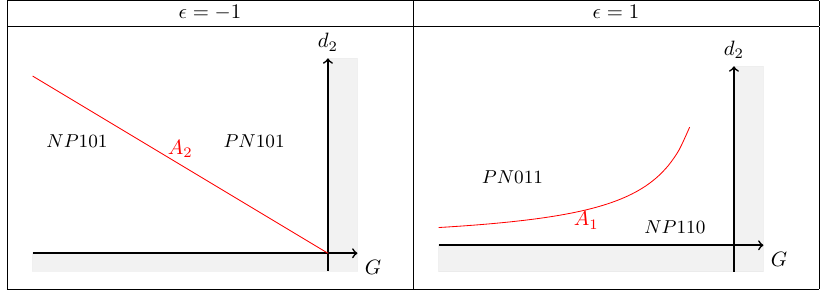}
  \caption{\label{tab:afocal_codim2}Afocal codim 2}
  \end{center}
  \end{figure}

  Let us remark that in this case, ${\mathcal{G}_{o}}$ depends only on $\mathbf{y}$ so that $\mathcal{G}_{oo}=\mathcal{G}_{o}\circ \pi$.
  \paragraph{Focal case }
     We recall that $f\in \{-1,1\}$, we set $\mathbf{y}=(d_1,d_2,d_3)$ and $\mathbf{x}=(\Omega_1,\Omega_2)$. 
   We set $\Goo(\mathbf{y})=(d_1<0)\land (d_2>0)\land (d_3<0)$.
    We consider the ideal $I=\langle g_1,g_2\rangle$ on $\mathbb{K}[\mathbf{y},\mathbf{x}]$ with $\mathbb{K}=\mathbb{Q}(f)$.
   A Groebner basis of $I$ for $\textrm{revlex}(\mathbf{x})\succ\textrm{revlex}(\mathbf{y})$ is $\{d_3\wt{g_1},d_3\wt{g_2},\wt{g_3}\}$ where
   \begin{equation}
   \begin{aligned}
   \wt{g_1}=A_1\Omega_1^2+B_1\Omega_1+C_1,\quad
    \wt{g_2}= A_2\Omega_2+B_2,\quad 
    \wt{g_3}=\Omega_1\Omega_2f+d_3,
   \end{aligned}
   \label{eq:triangularFocalCodim2}
   \end{equation}
   with 
   \begin{align*}
       A_1=f(d_1f+d_2d_3),\quad 
       B_1=f(-d_1d_2+2d_1d_3-2d_2d_3),\quad 
       C_1=d_3C_1'\\
       C_1'=d_1d_3+d_2f,\quad
       A_2=C_1'f,\quad
       B_2=u_2\Omega_1+v_2,\quad u_2=A_1,\quad v_2=B_1.
   \end{align*}
   We deduce that $\tilde{\mathbf{f}}\overset{\text{def}}{=}(\wt{g_1},\wt{g_2},\wt{g_3})$ is of the form \eqref{eq:triangularQuadraticSystemCodim23} and that \cref{hyp:codim} is verified with $\n=1$. 
   We set $\wt{I}=\langle \tilde{\mathbf{f}}\rangle$.
   We easily check \cref{hyp:AkC0} with $E_{oo}\supset E$ and deduce that \cref{theo:xi1epsContinuousThroughA1epsB1} holds.
  The discriminant defining $F_o=\set{\Delta\geq0\land\mathcal{G}_{oo}}$ is given by 
   \[
    \Delta =-fd_1d_2\left( -fd_1d_2+4d_3((f+d_3)^2+fd_1-fd_2)\right).
       \]
       Let us remark that for $f=-1$, $\Delta>0$ on $F_{oo}$.
%   The set $F_\epsilon$ is given by 
%    \begin{align*}
%    F_\epsilon &= F_o\backslash\mathcal{W}_\infty^{(\epsilon)}\\
%    \mathcal{W}_1^{(\epsilon)}&=\mathcal{W}_{A_1}\cap \{\epsilon B_1\leq 0\}\\
%        \mathcal{W}_2^{(\epsilon)}&=\mathcal{W}_{A_2}\cap\{\epsilon B_1\geq 0\}.
%    \end{align*}
   By \eqref{eq:AssumptionV_QVerified} and \eqref{eq:Q_kFocal}, we get that 
   \[
   \mathcal{V}_{Q_k}\cap E_{oo} =\mathbf{V}( q_k\ \alpha_kx_1+\beta_k,\ \tilde{f}_2,\tilde{f}_3)\cap E_{oo},
   \]
        with $q_k$ obtained by computing  $G_k\cap \mathbb{Q}[\mathbf{y}]$ with $G_k$ a Groebner basis of $I_k=(\wt{I}+\langle Q_k\rangle) \cap\mathbb{Q}[\mathbf{y}]$ for $\textrm{revlex}(\mathbf{x})\succ\textrm{revlex}(\mathbf{y})$:
    \begin{align*}
            G_1\cap \mathbb{Q}[\mathbf{y}] &=d_1(-d_2f + d_3^2 + 2d_3f + f^2):=d_1q_1\\
            G_2\cap \mathbb{Q}[\mathbf{y}]&=d_1d_2(d_1f - d_2f + d_3^2 + 2d_3f + f^2):=d_1d_2q_2\\
            G_3\cap \mathbb{Q}[\mathbf{y}]& =d_2d_3(d_1f + d_2^2 - 2d_2d_3 - 2d_2f + d_3^2 + 2d_3f + f^2):=d_2d_3q_3.
        \end{align*}
        Hence $\mathcal{V}_{Q_k}^{(\epsilon)}$ writes as stated in \cref{le:expressionVQk}
        % \begin{align*}
        %     \mathcal{V}_{Q_k}^{(\epsilon)}&= \{ \mathbf{y}\in F_\epsilon,\ A_2(\mathbf{y})x_1+B_2(\mathbf{y},x_1)=0,\ \alpha_k(\mathbf{y})x_1+\beta_k(\mathbf{y})=0,\ q_k(\mathbf{y})=0,\ \epsilon s_k(\mathbf{y})\leq 0\}
        % \end{align*}
       with       
        \[    
        \begin{array}{c|c|c}
         \begin{aligned}
            \alpha_1=1,\ \beta_1=-1\\
            \alpha_2=f(d_1-d_2),\ \beta_2=d_1d_3+d_2f\\
            \alpha_3=f,\ \beta_3=d_3-d_2
        \end{aligned}&
        \begin{aligned}
            q_1&=-fd_2+(f+d_3)^2\\
            q_2&=f(d_1-d_2)+(f+d_3)^2\\
            q_3&=fd_1+(f-d_2+d_3)^2\\
        \end{aligned}&
        \begin{aligned}
            s_1&=- \left(d_{3} - f\right) \left(d_{3} + f\right)\\
            s_2&=\left(-d_{2} f + (d_{3}+ f)^2\right)\\
            s_3&=(d_2-3d_3-f)(d_2-d_3-f)
        \end{aligned}
        \end{array}
        \]
        where $s_k$ is a representant of $-\alpha_k(-2A_1\beta_k+\alpha_kB_1)$ in $\mathbb{K}[\mathbf{y}]/\langle q_k\rangle $ up to  positive  factors as  terms of the form $(-1)^kd_k$. 
        We deduce that \cref{hyp:NoralFormQk} is verified and we get that \cref{theo:CCdeGepsetCCdeE} holds true.
        For $f=1$, we obtain that  $G_\epsilon = F_o\backslash\cup_k\mathcal{W}_{Q_k}^{(\epsilon)}\quad\text{with}\quad 
        \mathcal{W}_{Q_k}^{(\epsilon)} = \{q_k(\mathbf{y})=0\land \epsilon s_k\leq 0\}$.
        The pairwise intersections of the sets $\mathcal{W}_{Q_k}^{(\epsilon)}$ for $k\in \llbracket 1,3\rrbracket$ are easily obtained by using Groebner basis computation (see \cref{tab:focal_codim2_fpos}).
        For $f=-1$, we get that $G_\epsilon = F_o\backslash\cup_k\mathcal{W}_k^{(\epsilon)}$ with $\mathcal{W}_1^{(\epsilon)} = F_o\cap \set{(A_1=0)\land (\epsilon B_1\leq 0)}$ that is $\mathcal{W}_1^{(1)}=\emptyset$ and $\mathcal{W}_1^{(-1)}=\mathcal{W}_{A_1}$, and  $\mathcal{W}_2^{(\epsilon)}=F_o\cap \set{(A_2=0)\land (\epsilon B_1\geq 0)}$ that is $\mathcal{W}_2^{(1)}=\mathcal{W}_{A_2}$ and  $\mathcal{W}_2^{(-1)}=\emptyset$.
By sampling the connected component of $\{\mathbf{y}=(d_1,d_2,d_3)\in\mathbb{R}^3\ :\  \Delta(\mathbf{y})>0\land d_1<0\land d_2>0\land d_3>0\land (A_1A_2q_1q_2q_3)(\mathbf{y})\neq 0\}$ for $f\in \{-1,1\}$, we get a list of topological invariant name / point / branch corresponding to $L_o^{(\epsilon)}$ used in \cref{alg:computeGeps} and associated to $\CC^{(\epsilon)}$ for $f=1$:
\begin{align*}
 (\texttt{PP010}, (-10, 2, -2),\epsilon=-1),
 (\texttt{PP011}, (-5, 2, -2),\epsilon=-1),\\
 (\texttt{PP110}, (-17/16, 3/8, -3/8),\epsilon=-1),
 (\texttt{PP001}, (-31/64, 1/2, -2),\epsilon\in \{-1,1\}),\\
 (\texttt{PP100}, (-3/1024, 21/32, -3/16),\epsilon\in \{-1,1\}),
 (\texttt{PP101}, (-5/16, 1/4, -3/16),\epsilon\in \{-1,1\})
 \end{align*}
and for $f=-1$: 
\begin{align*}
    (\texttt{NP101}, (-1, 1/2, -3),\epsilon=-1), \ (\texttt{PN011}, (-1, 1/2, -3),\epsilon=1), \\
    (\texttt{PN101}, (-1, 2, -1/4),\epsilon=-1),\ (\texttt{NP110}, (-1, 1/2, -3),\epsilon=1).
\end{align*}
 Hence, by setting $h_{\epsilon}(\mathbf{y})=(\mathbf{y},\xi^{(\epsilon)}(\mathbf{y}))$, we get  the following sets composing $\CC^{(\epsilon)}$ denoted uniquely by their topological invariant.\\
 \noindent \textbf{(i) Case $\bm{f=1,\ \epsilon=-1}$}
    \begin{align*}
            C_1^{(-1)}=\texttt{PP011}^{(-1)}&=h_{-1}\left(\{ d_1<0,\ d_2>0,\ d_3<0,\ \Delta\geq 0,\ q_2(\mathbf{y})<0,\ q_3(\mathbf{y})>0 \}\right)\\
            C_2^{(-1)}=\texttt{PP010}^{(-1)}&=h_{-1}\left(\{ d_1<0,\ d_2>0,\ d_3\leq -1,\ q_3(\mathbf{y})<0 \}\right)\\
            &\cup h_{-1}\left(\{ d_1<0,\  -1\leq d_3<0,\ (1+d_3)^3<d_2<(1+d_3)\}\right)\\
            C_3^{(-1)}=\texttt{PP110}^{(-1)}&=h_{-1}\left(\{ d_1<0,\ d_2>0,\ d_3<0,\ \Delta\geq 0,\ q_1(\mathbf{y})>0,\ q_2(\mathbf{y})<0 \}\right)\\
            %&=\{(\mathbf{z},\mathbf{y})\in \mathbb{R}^5,\ \mathbf{y}\in \mathcal{W}_{oo},\ q_1(\mathbf{y})>0,\ q_2(\mathbf{y})<0,\ \Omega_1=\frac{-B-\sqrt{\Delta}}{2A}, \Omega_2=\frac{-d_3}{\Omega_1f}\}\\
            C_4^{(-1)}=\texttt{PP101}^{(-1)}&=h_{-1}\left(\{ d_1<0,\ d_2>0,\ -\frac{1}{3}\leq d_3<0,\ \Delta\geq 0,\ q_3(\mathbf{y})>0,\
            \right.\\
            &\quad \left. -(1+d_3)^2\leq d_1\leq -4d_3^2,\ d_2\leq (1+3d_3) \}\right)\\
             C_5^{(-1)}=\texttt{PP100}^{(-1)}&=h_{-1}\left( \{ d_1<0,\ d_2>0,\ -1\leq d_3<0,\ q_2(\mathbf{y})>0,\ -4d_3^2\leq d_1 ,\ \Delta\geq 0 \}\right)\\
             &\cup h_{-1}\left( \{ d_1<0,\ d_2>0,\ -1\leq d_3<0,\ q_2(\mathbf{y})>0,\ -(1+d_3)^2\leq d_1\leq -4d_3^2,\ q_3(\mathbf{y})< 0 \}\right) \\
             C_6^{(-1)}=\texttt{PP001}^{(-1)}&=h_{-1}\left(\{ d_1<0,\ d_2>0,\ d_3\leq-1,\ \Delta\geq 0,\ q_2(\mathbf{y})>0 \}\right)
        \end{align*}
\noindent \textbf{(ii) Case $\bm{f=1,\ \epsilon=1}$}
    \begin{align*}
        C_4^{(1)}=\texttt{PP101}^{(1)}&=h_{1}\left(\{ d_1<0,\ d_2>0,\ d_3\leq -1,\ \Delta\geq 0,\  q_1(\mathbf{y})<0 \}\right)\\
            &\cup h_{1}\left( \{d_1<0,\ d_2>0,\ -1\leq d_3<0,\ \Delta\geq 0,\  q_3(\mathbf{y})<0) \}\right)\\
            &\cup h_{1}\left( \{d_1<0,\ d_2>0,\ -\frac{1}{3}\leq d_3<0,\ \Delta\geq 0 \}\right)\\
            C_5^{(1)}= \texttt{PP100}^{(1)}&=h_{1}\left(\{ d_1<0,\ -1\leq d_3<0,\ 1+3d_3\leq d_2\leq 1+d_3,\ \Delta\geq 0,\ \ q_3(\mathbf{y})>0\}\right)\\
            C_6^{(1)}=\texttt{PP001}^{(1)}&=h_{1}\left(\{ d_1<0,\ d_2>0,\ d_3\leq-1,\ \Delta\geq 0,\ q_1(\mathbf{y})>0 \}\right)
    \end{align*}
\noindent \textbf{(iii) Case $\bm{f=-1,\ \epsilon=-1}$}
\begin{align*}
            D_1^{(-1)}=\texttt{PN101}^{(-1)}&=h_{-1}\left(\{ d_1<0,\ d_2>0,\ d_3<0,\ A_2(\mathbf{y})<0 \}\right)\\
            D_2^{(-1)}=\texttt{NP101}^{(-1)}&=h_{-1}\left(\{ d_1<0,\ d_2>0,\ d_3<0,\ A_2(\mathbf{y})>0 \}\right)
        \end{align*}
        \noindent \textbf{(iv) Case $\bm{f=-1,\ \epsilon=1}$}
        \begin{align*}
            D_3^{(1)}=\texttt{PN011}^{(1)}&=h_{1}\left(\{ d_1<0,\ d_2>0,\ d_3<0,\ A_1(\mathbf{y})>0\}\right)\\
            D_4^{(1)}=\texttt{NP110}^{(1)}&=h_{1}\left(\{ d_1<0,\ d_2>0,\ d_3<0,\ A_1(\mathbf{y})<0 \}\right)
        \end{align*}
        The sets \( C_4^{(\pm 1)} \), \( C_5^{(\pm 1)} \), and \( C_6^{(\pm 1)} \), each associated with the respective topological invariants \texttt{PP101}, \texttt{PP100}, and \texttt{PP001}, are combined through \( E_\mathcal{H} \) into three connected sets: \( C_4 \), \( C_5 \), and \( C_6 \).
        We conclude that the resulting sets, outputs of Algorithm~\ref{alg:computeGeps}, represent the connected components of \( E \).  
        Ultimately, we obtain a list of connected sets, each associated with distinct topological invariants for the cases \( f \in \{-1,1\} \), confirming that \( \mathcal{S} \) as defined in \cref{def:nomenclature} is exact.
        These topological invariants are summarized in \cref{tab:summaryclassif3MA} (codimension 2). Illustrations are provided in \cref{tab:focal_codim2_fpos} and \cref{tab:focal_codim2_fneg}.
  \paragraph{Afocal system}
   We recall that $d_1=-1$, we set $\mathbf{x}=(\Omega_1,\Omega_2)$ and $\mathbf{y}=(G,d_2)$.
 We set $\Goo(\mathbf{y})=(G\neq0)\land (d_2>0)$.
  We consider the ideal $I=\langle h_1,h_2\rangle$ on $\mathbb{K}[\mathbf{y},\mathbf{x}]$ with $\mathbb{K}=\mathbb{Q}(d_1)$. 
 A Groebner basis  of $I$ for $\textrm{revlex}(\mathbf{x})\succ\textrm{revlex}(\mathbf{y})$ is $\{\wt{h_1},\wt{h_2},\wt{h_3}\}$ with :
   \begin{align*}
    \wt{h_1}= A_1\Omega_1^2+B_1\Omega_1+C_1,\quad
    \wt{h_2}=A_2\Omega_2+B_2,\quad
    \wt{h_3}=\Omega_1\Omega_2-G,
\end{align*}
and 
\begin{align*}
 A_1=-Gd_2 + d_1,\quad
    B_1=2G(d_2-d_1 ),\quad
    C_1=G A_2\\
    A_2=Gd_1 - d_2,\quad
    B_2=u_2\Omega_1 +v_2 ,\quad u_2=A_1,\quad v_2=B_1.
\end{align*}
Hence, $\tilde{\mathbf{f}}\overset{\text{def}}{=}(\wt{h_1},\wt{h_2},\wt{h_3})$ takes the form \eqref{eq:triangularQuadraticSystemCodim23} and \cref{hyp:codim} holds with $\n=1$.
 We easily check \cref{hyp:AkC0} with $E_{oo}\supset E$ and deduce that \cref{theo:xi1epsContinuousThroughA1epsB1} holds.
The set $F_o$ is given by $F_o=\set{(\Delta\geq 0)\land \Goo}$ with $\Delta=4Gd_1d_2(G - 1)^2$
    and verifies $\sign(\Delta)=-\sign(G)$ which leads to 
    $F_o = \{(d_2,G)\in \mathbb{R}^2,\ d_2>0,\ G<0\}$.
   Let us remark that  $\sign(B_1)=\sign(G)$ which is negative on $F_o$, so that
    $\mathcal{W}_1^{(1)}=\mathcal{W}_{A_1}$, $\mathcal{W}_1^{(-1)}=\emptyset$,
       $\mathcal{W}_2^{(-1)}=\mathcal{W}_{A_2}$ and $\mathcal{W}_2^{(1)}=\emptyset$.
  By  \eqref{eq:AssumptionV_QVerified} and \eqref{eq:Q_kAFocal} we get that $\mathcal{V}_{Q_k}\cap E_{oo} =\mathbf{V}( q_k\ \alpha_kx_1+\beta_k,\ \tilde{f_2},\tilde{f}_3)\cap E_{oo}$ and
        % \begin{align*}
        %     \mathcal{V}_{Q_k}^{(\epsilon)}= \{ \mathbf{y}\in F_\epsilon,\ A_2(\mathbf{y})x_2+B_2(\mathbf{y},x_1)=0,
        %       \alpha_k(\mathbf{y})x_1+\beta_k(\mathbf{y})=0,\ q_k(\mathbf{y})=0,\ \epsilon s_k(\mathbf{y})\leq 0\}
        % \end{align*}
        $\mathcal{V}_{Q_k}^{(\epsilon)}$ writes as stated in \cref{le:expressionVQk} with 
        \[    
        \begin{array}{c|c|c}
         \begin{aligned}
            \alpha_1=1,\ \beta_1=-1\\
            \alpha_2=(d_1-d_2),\ \beta_2=d_2-Gd_1\\
            \alpha_3=-1,\ \beta_3=G
        \end{aligned}&
        \begin{aligned}
            q_1&=q_2=q_3=G-1\\
        \end{aligned}&
        \begin{aligned}
            s_1&=s_2=s_3=0\\
        \end{aligned}
        \end{array}
        \]
        showing that \cref{hyp:NoralFormQk} is verified and \cref{theo:CCdeGepsetCCdeE} holds true.
         As for all $k\in \llbracket 1,3\rrbracket$, $q_k$ does not cancel out on $F_{o}\subset \set{G<0}$,   we deduce that $\mathcal{W}_{Q_k}^{(\epsilon)}=\emptyset$ and 
        \begin{align*}
            G_\epsilon &= F_\epsilon
            =\left\lbrace\begin{aligned}
                &F_o\backslash \mathcal{W}_{A_1},\ \textrm{ for } \epsilon=1\\
                &F_o\backslash \mathcal{W}_{A_2},\ \textrm{ for } \epsilon=-1\\
            \end{aligned}\right.
        \end{align*}
        By sampling the connected components of $\{(G,d_2)\in \mathbb{R}^2, d_2>0,\ G<0,\ A_1A_2\neq 0\}$ we get  $L=[(-2,1),(-1/2,1)]$ associated to the following list of topological invariant name / point / branch:
        \begin{align*}
            (\texttt{NP101},(-2,1),\epsilon=-1),(\texttt{PN101},(-1/2,1),\epsilon=-1),
            (\texttt{PN011},(-2,1),\epsilon=1),(\texttt{NP110},(-1/2,1),\epsilon=1)
        \end{align*}
        Hence, by setting $h_{\epsilon}(\mathbf{y})=(\mathbf{y},\xi^{(\epsilon)}(\mathbf{y}))$, the list $\CC^{(\epsilon)}$ is  composed  by the following sets:
        \begin{itemize}
            \item[] \textbf{(i) Case  $\bm{\epsilon=-1}$}
           \begin{align*}
               C_1^{(-1)}=\texttt{NP101}^{(-1)}&=h_{-1}\left(\{ G<0,\ d_2>0,\ A_2(\mathbf{y})>0\}\right)\\
               C_2^{(-1)}=\texttt{PN101}^{(-1)}&=h_{-1}\left(\{ G<0,\ d_2>0,\ A_2(\mathbf{y})<0\}\right)
           \end{align*}
           \item[]\textbf{(ii) Case  $\bm{\epsilon=1}$}
           \begin{align*}
            C_3^{(1)}=\texttt{PN011}^{(1)}&=h_{1}\left(\{ G<0,\ d_2>0,\ A_1(\mathbf{y})>0\}\right)\\
            C_4^{(1)}=\texttt{NP110}^{(1)}&=h_{1}\left(\{ G<0,\ d_2>0,\ A_1(\mathbf{y})<0\}\right)
           \end{align*}
        \end{itemize}
        Since  $\mathcal{W}_\mathcal{H}=\{G=0\}\cap F_{oo}=\emptyset$, this means that no merging is possible through $E_\mathcal{H}$ (see \cref{coro:ccIntersectInWh}), 
        we deduce that the above sets are the outputs of \cref{alg:computeGeps} and   represent  the connected components of $E$,
        each associated with distinct topological invariants showing again that $\mathcal{S}$ as defined in \cref{def:nomenclature} is exact. 
An illustration is given in \cref{tab:afocal_codim2} and a summary of the topological invariant names is given in \cref{tab:summaryclassif3MA} (codimension 2).
  \subsubsection{Systems of codimension $n=3$}
  Let us remark that in this case ${\mathcal{G}_{oo}}\neq  \mathcal{G}_{o}\circ \pi$.
  
\begin{figure}[htbp]
    \begin{center}
  \includegraphics[width=\textwidth]{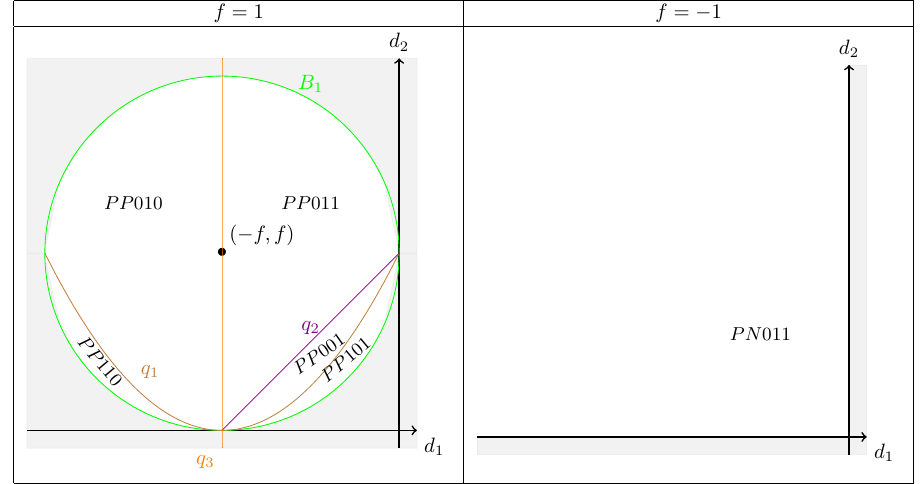}
  \caption{\label{tab:focal_codim_3}Focal codim 3, $f=\pm 1$}
  \end{center}
  \end{figure}

  \paragraph{Focal system}
  We set $\mathbf{x}=(d_3,\Omega_1,\Omega_2)$ and $\mathbf{y} = (d_1,d_2)$ and we consider the ideal $I=\langle g_1,g_2,g_3\rangle$ on $\mathbb{K}[\mathbf{y},\mathbf{x}]$ with $\mathbb{K}=\mathbb{Q}(f)$.
We set $\Goo(\mathbf{y})=(d_1<0)\land (d_2>0)$ and $\mathcal{H}_o(\mathbf{y},\mathbf{x})=(d_3<0)$ and $\Go(\mathbf{y},\mathbf{x})=\Goo(\mathbf{y})\land \mathcal{H}_o(\mathbf{y},\mathbf{x})$.
A Groebner basis of $I $ for $\textrm{revlex}(\mathbf{x})\succ\textrm{revlex}(\mathbf{y})$ in $\mathbb{K}[\mathbf{y},\mathbf{x}]$ gives  $\{ \wh{g_1},...,\wh{g_7}\}$. 
% \begin{align*}
%     \wh{g_1}&=d_{1}^{2} d_{3}^{2} f + d_{1} d_{2} d_{3}^{3} + 2 d_{1} d_{3}^{2} f^{2} + d_{2}^{2} d_{3}^{2} f - 2 d_{2} d_{3}^{2} f^{2} + d_{3}^{2} f^{3}
% \\
%     \wh{g_2}&=\Omega_{1} d_{3} - \frac{d_{1} d_{3}^{2}}{f^{2}} - \frac{d_{2} d_{3}}{f}\\
%     \wh{g_3}&=- \frac{\Omega_{2} d_{1}^{2} d_{2}^{2}}{2 f} - \Omega_{2} d_{1} d_{2}^{2} + \Omega_{2} d_{2}^{3} - \frac{\Omega_{2} d_{2}^{2} f}{2} - \frac{d_{1}^{2} d_{2} d_{3}}{2 f} - \frac{d_{1} d_{2}^{2} d_{3}^{2}}{2 f^{2}} - d_{1} d_{2} d_{3} + d_{2}^{2} d_{3} - \frac{d_{2} d_{3} f}{2}
% \\
%     \wh{g_4}&=\Omega_{1} \Omega_{2} + \frac{d_{3}}{f}\\
%     \wh{g_5}&=\Omega_{2} d_{1} d_{3} + \Omega_{2} d_{2} f + d_{3} f
% \\
%     \wh{g_6}&=\frac{\Omega_{2} d_{1} d_{2}^{2}}{2} + \Omega_{2} d_{2}^{2} d_{3} + \Omega_{2} d_{2}^{2} f - \frac{\Omega_{2} d_{2} d_{3} f}{2} + \frac{d_{1} d_{2} d_{3}}{2} + \frac{d_{2}^{2} d_{3}^{2}}{2 f} + d_{2} d_{3} f
% \\
%     \wh{g_7}&=- \frac{\Omega_{2} d_{2}^{2} f}{2} + \Omega_{2} d_{2} d_{3}^{2} + \Omega_{2} d_{2} d_{3} f - \frac{\Omega_{2} d_{3}^{2} f}{2} + \frac{d_{1} d_{3}^{2}}{2} + \frac{d_{2} d_{3}^{3}}{2 f} - \frac{d_{2} d_{3} f}{2} + d_{3}^{2} f
% \end{align*}
We can show that $d_2\wh{g_5},fd_1^2\wh{g_6},fd_1^2d_2\wh{g_7}\in \langle \wt{g_1},...\wt{g_4}\rangle$ and
%  and more precisely:
% \begin{align*}
% d_2\wh{g_5}&=\Omega_2(A_1d_3+d_2^2f)+d_3d_2f\\
% &=\Omega_2(\wt{g_1}-B_1+d_2^2f)+d_3d_2f\\
% &=\Omega_2\wt{g_1}+\Omega_2(-fA_3)+d_3d_2f\\
% &=\Omega_2\wt{g_1}+f\wt{g_3}\\
% fd_1^2\wh{g_6}&=(\Omega_2(2d_2f-f^2)+d_2d_3)\wt{g_1}+(d_2-f)^2f\wt{g_3}\\
% 2d_2\wh{g_7}&=2d_3\wh{g_6}-\Omega_2d_2(d_1d_2d_3) - \Omega_2d_2^3f - d_2^2d_3f\\
% &=2d_3\wh{g_6}-\Omega_2d_2(\wt{g_1}-B_1+d_2^2f)  - d_2^2d_3f\\
% &=2d_3\wh{g_6}-\Omega_2d_2(\wt{g_1}-fA_3)  - d_2^2d_3f\\
% &=2d_3\wh{g_6}-\Omega_2d_2\wt{g_1}-fd_2\wt{g_3}\\
% \end{align*}
 $\mathbf{V}(\wh{g_1},\wh{g_2},\wh{g_3},\wh{g_4})\cap E_{oo}= \mathbf{V}(\wt{g_1},...\wt{g_4})\cap E_{oo}$  with
\begin{align*}
    \wt{g_1}= A_1d_3+B_1,\quad
   \wt{g_2}= A_2\Omega_1+B_2,\quad 
   \wt{g_3}=A_3\Omega_2+B_3,\quad 
    \wt{g_4}=\Omega_1\Omega_2f+d_3,
\end{align*}
where
\[
\begin{array}{ccc}
A_1 = d_1d_2,&B_1=f((d_1+f)^2 +(d_2-f)^2-f^2)=f(A_3+d_2^2),\\
 A_2=f^2,& B_2=-d_1d_3 - d_2f,\\
A_3 = (d_1+f)^2-2d_2f,&B_3=-d_2d_3,\\
\end{array}
\]
% \begin{align*}
%     \wt{g_1}=\frac{\wh{g_1}}{d_3^2} = A_1d_3+B_1,\quad
%    \wt{g_2}=\frac{f^2\wh{g_2}}{d_3} = A_2\Omega_1+B_2\\
%    \wt{g_3}=\frac{2f}{d_2^2} \wh{g_3}+\frac{1}{d_3  d_2f  }\wh{g_1}=A_3\Omega_2+B_3,\quad 
%     \wt{g_4}=\wh{g_4}=\Omega_1\Omega_2f+d_3
% \end{align*}
and eventually that $\tilde{\mathbf{f}}\overset{\text{def}}{=}(\tilde{g_1},\dots,\tilde{g_4})$ is of the form \eqref{eq:triangularQuadraticSystemCodim23} with $\tilde{f_1}$
 having a null coefficient in front of $x_1^2$.
 Results of \cref{sec:triangular_system} easily adapt to this case:  there is only one branch solution $\xi\in C^0(F,\mathbb{R}^3)$ given by
 $\xi(\mathbf{y})=\left(-\frac{B_k(\mathbf{y})}{A_k(\mathbf{y})}\right)_{1\leq k\leq 3}$ where  $F\subset\mathbb{R}^2$ is defined here after.
 \cref{hyp:codim} is easily verified with 
\begin{align*}
   \n=1,\quad u_2=-d_1,\quad v_2=-d_2f,\quad u_3=-d_2,\quad v_3= 0.
\end{align*}
We remark that $\set{\mathcal{G}_o\land \left(\tilde{f_1}=0\right)}=\set{\mathcal{G}_{oo}\land (B_1<0)\land  (\tilde{f_1}=0)}$ so that  we set\footnote{For $f=1$, $F_o$ is the set of $(d_1,d_2)$  lying in the interior of the disk centered at $(-f,f)$ and of radius $f$. For $f=-1$ it is the quadrant where $d_1<0$ and $d_2>0$.}:
\[
    F_o = \set{ \mathcal{G}_{oo}\land (B_1<0)}.
\]
We remark that $\mathcal{W}_{A_1}=\mathcal{W}_{A_2}=\mathcal{W}_{A_3}=\emptyset$, hence $\mathcal{W}_\infty = \emptyset$  and we set $F = F_o$.
 We easily check \cref{hyp:AkC0} by withdrawing the conditions on $p_k$ and replacing $A_1$ and $B_1$ by respectively $B_1$ and $C_1$ and considering as supset   $E_{oo}\supset E$ so that \cref{theo:xi1epsContinuousThroughA1epsB1} applies and adapts as $E \cong \pi(E)$.
% \begin{align*}
%     E \cong \pi(E)%,\quad E_3 = \Gamma(\xi_{|\mathcal{W}_{A_3}}).
% \end{align*}
\noindent Hence, by checking \cref{hyp:NoralFormQk}, we get $E=E_o\backslash\cup_{k}(\mathcal{V}_{Q_k}\cap E_o)$
       with  $
            \mathcal{V}_{Q_k}\cap E_o= \{(\mathbf{y},\mathbf{x})\in \mathbf{V}_\mathbb{R}(\ftilde),\ \mathbf{y}\in F,\ q_k(\mathbf{y})=0\}$      
             with %$q_k$ such that $\langle q_k \rangle =(\langle \ftilde\rangle +\langle \mathcal{Q}_k\rangle)\cap \mathbb{Q}[\mathbf{y}]$:
        \begin{align*}   
            q_1=(d_1+f)^2 - d_2f, \quad 
            q_2=(-d_1 + d_2 - f),\quad
            q_3=d_1 + f,
        \end{align*}
        and $E = \set{(\wt{\mathbf{f}}=0)\land\mathcal{G}_{oo}\land(B_1<0)\land (q_1q_2q_3\neq0)}$.
Then following \cref{alg:computeGeps}, we sample the connected components of $\{\mathbf{y}=(d_1,d_2)\in\mathbb{R}^2,\ d_1<0,\ d_2>0,\ B_1(\mathbf{y})<0,\ (q_1q_2q_3)(\mathbf{y})\neq 0\}$, and we get the following list of pairs 
topological invariant name / point by set composing the set $\CC$: 
\begin{itemize}
    \item[]\textbf{(i) case $\bm{f=1}$}
\begin{align*}
    (\texttt{PP110},(-3/2, 3/16)),
    (\texttt{PP010},   (-3/2, 1)),
    % (\texttt{PP010},  [-3/2, 3/2]),
      (\texttt{PP101}, (-3/4, 3/64)),\\
      (\texttt{PP001}, (-3/4, 5/32)),
      (\texttt{PP011},(-3/4, 1)),
     % (\texttt{PP011}[-3/4, 3/2])
\end{align*}
    \item []\textbf{(ii) case $\bm{f=-1}$} : $(\texttt{PN011},(-1,1))$
\end{itemize}
By defining  $h(\mathbf{y})=(\mathbf{y},\xi(\mathbf{y}))$ and by applying \cref{coro:ccIntersectInWh} and \cref{theo:CCdeGepsetCCdeE}, the following list of sets output of \cref{alg:computeGeps} are the connected components of $E$:
\begin{itemize}
    \item \textbf{(i) case $\bm{f=1}$}
    \begin{align*}
        C_1&=\texttt{PP110}=h\left(\{  d_1<-f,\ d_2>0,\ B_1(\mathbf{y})<0,\ q_1(\mathbf{y})>0\}\right)\\
        C_2&=\texttt{PP010}=h\left(\{  d_1<-f\ d_2>0,\ B_1(\mathbf{y})<0,\ q_1(\mathbf{y})<0\}\right)\\
        %C_3&=\texttt{PP010}^{(b)}=h\left(\{ d_1<0,\ d_2>0,\ B_1(\mathbf{y})<0,\ d_2>f,\ d_1<-f\}\right)\\
        %C_3&=\texttt{PP011}^{(a)}=h\left(\{ d_1<0,\ d_2>0,\ B_1(\mathbf{y})<0,\ d_2>f,\ d_1>-f\}\right)\\
        C_3&=\texttt{PP011}=h\left(\{ -f<d_1<0,\ d_2>0,\ B_1(\mathbf{y})<0,\ q_2(\mathbf{y})>0\}\right)\\
        C_4&=\texttt{PP001}=h\left(\{ -f<d_1<0,\ d_2>0,\ B_1(\mathbf{y})<0,\ q_2(\mathbf{y})<0,\  q_1(\mathbf{y})<0\}\right)\\
        C_5&=\texttt{PP101}=h\left(\{ -f<d_1<0,\ d_2>0,\ B_1(\mathbf{y})<0,\ q_1(\mathbf{y})>0\}\right)\\
    \end{align*}
    \item[]\textbf{(i) case $\bm{f=-1}$}
    \begin{align*}
        D_1&=\texttt{PN011}=\set{\mathcal{G}_{oo}}
    \end{align*} 
\end{itemize} 
An illustration is given in \cref{tab:focal_codim_3} and a summary of the topological invariants  in \cref{tab:summaryclassif3MA} (codimension 3).
Let us remark that all the connected components are associated to a different topologcial invariant name, meaning that $\mathcal{S}$ in \cref{def:nomenclature} is exact.
%Let us remark that the components $C_2$ and $C_3$ (resp. $C_4$ and $C_5$), share the same topological invariant $\texttt{PP010}$ (resp. $\texttt{PP011}$) but are not in the same connected component.
%This demonstrates that the topological invariant $\mathcal{S}$ in \cref{def:nomenclature} is \textbf{not exact}.
% but are separated by the vanishing algebraic set associated to  $I+\langle\wt{Q_2}\rangle$ which is not in the set of admissible optical solutions.
%This shows that $\mathcal{S}$  is not maximal in the sense that two points in $E$ with the same topological invariant are not necessarily in the same connected component (see Remark~\ref{re:topologicalInvariantDef}). 

  \paragraph{Afocal system}
  
\begin{figure}[htbp]
    \begin{center}
  \includegraphics[width=\textwidth]{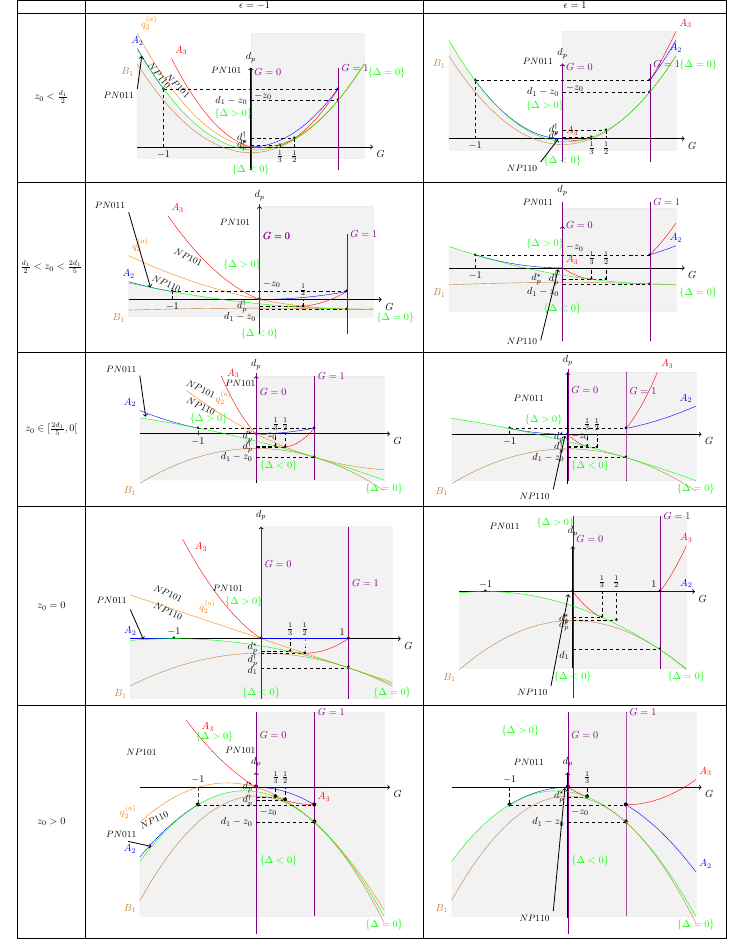}
  \caption{\label{tab:classifAfocalCodim3}Afocal of codim 3, $d_p^\star =\frac{-z_0+4d_1}{9}$ and  $d_p^\dagger=-G^2z_0+Gd_1$}
  \end{center}
  \end{figure}

  We recall that we can set $d_1=-1$ and we set $\mathbf{x}=(d_2,\Omega_1,\Omega_2)$ and  $\mathbf{y}=(G,z_0,d_p)$.
We set $\Goo(\mathbf{y})=(G\neq 0)$, $\mathcal{H}_o(\mathbf{y},\mathbf{x})=(d_2>0)$ and $\Go(\mathbf{y},\mathbf{x})=\Goo(\mathbf{y})\land \mathcal{H}_o(\mathbf{y},\mathbf{x})$.
We consider the ideal $I=\langle h_1,h_2,h_3\rangle$ on $\mathbb{K}[\mathbf{y},\mathbf{x}]$ with $\mathbb{K}=\mathbb{Q}(d_1)$. 
 A Groebner basis of $I$ for $\textrm{revlex}(\mathbf{x})\succ\textrm{revlex}(\mathbf{y})$ is  $\left\lbrace \wt{h_1},\wt{h_2},\wt{h_3},\wt{h_4}, \wt{h_5},\wt{h_6}\right\rbrace$.
%  with
% \begin{align*}
%     \wt{h_1}&=A_1d_2^2+B_1d_2+C_1\\
%     \wt{h_2}&=A_2\Omega_1+B_2\\
%     \wt{h_3}&=A_3\Omega_2+B_3\\
%     \wt{h_4}&=\Omega_1\Omega_2-G\\
%     \wt{h_5}&=- G^{2} z_{0} + G \Omega_{1} d_{2} + 2 G d_{1} - 2 G d_{2} - \Omega_{1} d_{1} - d_p\\
%     \wt{h_6}&=G^{2} z_{0} - G \Omega_{2} d_{1} + \Omega_{2} d_{2} + d_p
% \end{align*}
By remarking that $\Omega_1\wt{h_6},\Omega_2\wt{h_5}\in \langle \wt{h_1},\dots,\wt{h_4}\rangle$, we get $\mathbf{V}(h_1,h_2,h_3)\cap E_{oo}=\mathbf{V}(\wt{h_1},\dots,\wt{h_4})\cap E_{oo}$
  with
\begin{align*}
    \wt{h_1}=A_1d_2^2+B_1d_2+C_1,\quad
    \wt{h_2}=A_2\Omega_1+B_2,\quad
    \wt{h_3}=A_3\Omega_2+B_3,\quad
    \wt{h_4}=\Omega_1\Omega_2-G,
\end{align*}
where 
\[
\begin{array}{ccc}
    A_1= G^2,&
B_1= -G^3d_1  - Gd_1 + 2GA_2,&
C_1= (G^2z_0 - Gd_1 + d_p)^2\\
   & A_2= G^2z_0 + d_p, &
B_2= Gd_2 -G^2d_1 ,\\
&A_3= 2G^2d_1  - 2Gd_1 + A_2,&
B_3= - G^2d_2  + Gd_1 - 2GA_2. \\
\end{array}
\]
Hence, $\tilde{\mathbf{f}}\overset{\text{def}}{=}(\wt{h_1},\dots,\wt{h_4})$ takes the form \eqref{eq:triangularQuadraticSystemCodim23} and \cref{hyp:codim} is easily checked with
\begin{align*}
    \n=-G,\ u_2=G,\quad v_2=-G^2d_1,\quad u_3=-G^2,\quad v_3= G(-2A_3+d_1).
\end{align*}
The discriminant defining $F_o$ is given by
$\Delta = -d_1G^2(G-1)^2(-d_1(G+1)^2+4A_2)$
    whose the sign  on $F_{oo}$ is the same as the one of $-d_1(G+1)^2+4A_2$. Let us set
    \[
    \wt{F_o} = \{(G,z_0,d_p)\in \mathbb{R}^3,\ G\neq 0, \ -d_1(G+1)^2+4A_2\geq 0\}.   
    \]
    By noting that \( \Delta < B_1^2 \), we deduce that $\set{\Go\cap(\tilde{f_1}=0)}=\set{\Goo\cap(B_1<0)\cap(\tilde{f_1}=0)}$ so that we set  $F_o=\wt{F_o} \cap \{ B_1 < 0 \}$
     which, can also be written as $F_o = \wt{F_o} \cap \{ G < 0 \}=\set{(\Delta\geq 0)\land (G<0)}$.     
    We  notice that $\mathcal{W}_{A_1}=\emptyset$ and
    % we check that \eqref{eq:relationA1v2B1v2Continuitexi} are verified for respectively $(\wt{h_1},\wt{h_2})$ and $(\wt{h_1},\wt{h_3})$ with
    % \begin{itemize}
    %     \item $2A_1v_2-B_1u_2=u_2P_2+T_1A_2$ with  $P_2=-d_1G(G^2-1)$, $T_1=-2G^2$ and $\Delta =P_2^2+T_2A_2$ with  $T_2=-4G^2d_1(G - 1)^2$.
    %     \item $2A_1v_3-B_1u_3=u_3P_3+\wt{T_1}A_3$ with $P_3=-d_1G(G-1)(3G-1)$, $\wt{T_1}=-2G^3$ and $\Delta =P_3^2+T_2A_3$ with  $T_2$ as above.
    %     \item $\wt{E_\infty}=\emptyset$
    % \end{itemize}
    % Hence we get
    by  
    \cref{le:compatibility}, we get $p_2=-p_3=- G \left(G^{2} d_{1} + 2 G^{2} z_{0} - d_{1} + 2 dp\right)
$. Let $\overline{p_k}$ be a representant of $p_k$ in $\mathbb{Q}[\mathbf{y}]\backslash\langle A_k\rangle$,
we get $\mathcal{W}_k^{(\epsilon)}=\mathcal{W}_{A_k}\cap\{\epsilon \overline{p_k}\geq 0\}$.
We carefully check that $(\mathbf{y},\mathbf{x})\in E_o\land (A_k=0)\land (p_k=0)=\texttt{False}$ and deduce that \cref{hyp:AkC0} is verified.
          Then computing a Groebner basis of $\langle\tilde{\mathbf{f}}\rangle+\langle \mathcal{Q}_k\rangle$ for $\textrm{revlex}(\mathbf{x})\succ\textrm{revlex}(\mathbf{y})$ leads to $\mathcal{V}_{Q_k}\cap E_{oo}=\mathbf{V}(q_k,\alpha x_1+\beta_1,\tilde{f_2},\dots,\tilde{f_4})\cap E_{oo}$ and we deduce that $\mathcal{V}_{Q_k}^{(\epsilon)}$ expresses as  in \cref{le:expressionVQk}
    % \begin{align*}
    %     \mathcal{V}_{Q_k}^{(\epsilon)}= \{\mathbf{y}\in {F_\epsilon},\ f_2(\mathbf{y},\mathbf{x})=f_3(\mathbf{y},\mathbf{x})=0,
    %     & \quad \alpha_k(\mathbf{y})x_1+\beta_k(\mathbf{y})=0,\ q_k(\mathbf{y})=0,\ \epsilon s_k\leq 0\}
    % \end{align*}
   with   
    \[    
    \begin{array}{c|c}
     \begin{aligned}
        \alpha_1=1,\ \beta_1=-Gd_1 - Gd_p + Gz_0 + 2d_p\\
        \alpha_2=1,\ \beta_2=-(G - 2)(G^2z_0 - Gd_1 + d_p)\\
        \alpha_3=1,\ \beta_3=G^2z_0 - Gd_1 + d_p
    \end{aligned}&
    \begin{aligned}
        q_1&=(G-1)\\
        q_2&=(G-1)(G^2z_0+d_p-Gd_1)\\
        q_3&=q_2,\\
    \end{aligned}    
    \end{array}
    \]
    where $s_k$ is the reduction of $-\alpha_k(-2A_1\beta_k+\alpha_kB_1)$ by $q_k$ up to  positive  factors as  terms of the form $(-1)^kd_k$:
    \begin{align*}
        s_1=0,\quad s_2=s_3=-G.
    \end{align*}
    We deduce that \cref{hyp:NoralFormQk} is verified and \cref{theo:CCdeGepsetCCdeE} applies.
    Note that $q_1=(G-1)$ never cancels on $F_o$, hence  we set $q_2^{(a)}=(G^2z_0+d_p-Gd_1)$,
    % Hence, let $\ov{s_2}$ (resp. $\ov{s_3}$) be the reduction of $s_2$ (resp. $s_3$) by $q_2^{(a)}=q_3^{(a)}=\frac{q_2}{G-1}=A_2-Gd_1$ up to constant sign term:
    % \begin{align*}
    %     \ov{s_1}=0,\quad \ov{s_2}=\ov{s_3}=-G 
    % \end{align*}
     $\mathcal{W}_{Q_2}^{(\pm 1)}=\mathcal{W}_{Q_3}^{(\pm 1)}$ and $\mathcal{W}_{Q_1}^{(\pm 1)}=\emptyset$ and we deduce the expression of $G_\epsilon$:
    \begin{align*}
        G_\epsilon=F_\epsilon\backslash \mathcal{W}_{Q_2}^{(\epsilon)}\quad \text{with}\quad
        \mathcal{W}_{Q_2}^{(\epsilon)} = \{\mathbf{y}\in F_o\ :\ q_2^{(a)}(\mathbf{y})=0\land \epsilon s_2(\mathbf{y})\leq 0 \}.
    \end{align*}
    By sampling the connected components of $\{\mathbf{y}=(G,z_0,d_p)\in \mathbb{R}^3,\  G< 0,\ -d_1(G+1)^2+4A_2 >0,\ (A_2A_3q_2^{(a)})(\mathbf{y})\neq 0  \}$, we get $L=[(-2,-1/32,0),(-1/2,-1/8,0),(-2,1,0),(-2,4,0)]$ associated to the following list of topologcial invariant name / point / branch : 
    \begin{align*}
        &(\texttt{PN011},(-2,-1/32,0),\epsilon=-1),\ (\texttt{NP110}, (-1/2,-1/8,0),\epsilon=-1),\ (\texttt{NP101},(-2,1,0),\epsilon=-1),\\
        & (\texttt{PN101}, (-2,4,0),\epsilon=-1), \ (\texttt{PN011},(-2,-1/32,0),\epsilon=1),\ (\texttt{NP110}, (-2,-1/8,0),\epsilon=1)
    \end{align*}
    We now refer the reader to \cref{tab:classifAfocalCodim3}. % and calculus of Appendix~\ref{sec:appendixAfocalCodim3}.
    By recalling that $h_{\epsilon}(\mathbf{y})=(\mathbf{y},\xi^{(\epsilon)}(\mathbf{y}))$, the following  collection is obtained in the before last step of \cref{alg:computeGeps}:
    \begin{itemize}
        \item[]\textbf{(i) Case $\bm{\epsilon=-1}$}
    \begin{align*}
        C_1^{(-1)}=\texttt{PN011}^{(-1)}&=h_{-1}\left(\{A_2<0,\ G<-1,\  \Delta \geq 0,\ \}\right)\\
        C_2^{(-1)}=\texttt{NP110}^{(-1)}&=h_{-1}\left(\{A_2>0,\ G<-1,\ q_2^{(a)}> 0,\ \}\right)\\
        &\cup h_{-1}\left(\{\Delta\geq 0,\ -1\leq G<0,\  q_2^{(a)}< 0 \}\right)\\
        C_3^{(-1)}=\texttt{NP101}^{(-1)}&=h_{-1}\left(\{A_3<0,\ q_2^{(a)}>0,\}\right)\\
        C_4^{(-1)}=\texttt{PN101}^{(-1)}&=h_{-1}\left(\{A_3>0\}\right)\\
    \end{align*}
    \item[]\textbf{(ii) Case $\bm{\epsilon=1}$}
    \begin{align*}
        C_1^{(1)}=\texttt{PN011}^{(1)}&= h_{1}\left(\{A_2>0,\ -1\leq G<0, \}\right)\cup h_{1}\left(\{\Delta\geq 0,\ G<-1, \}\right)\\
        C_2^{(1)}=\texttt{NP110}^{(1)}&=h_{1}\left(\{A_2<0,\ G\leq -1,\ \Delta\geq 0\}\right).
    \end{align*}
\end{itemize}
The sets \( C_1^{(\pm 1)} \) and \( C_2^{(\pm 1)} \) each associated with the respective topological invariants \texttt{PN011} and \texttt{NP110}, are combined through \( E_\mathcal{H} \) into two connected sets: \( C_1 \) and \( C_2 \).
We conclude that the resulting sets, outputs of \cref{alg:computeGeps}, represent the connected components of \( E \).  
Ultimately, we obtain a list of connected sets, each associated with distinct topological invariants, confirming that \( \mathcal{S} \) as defined in \cref{def:nomenclature} is exact.
A summary of the topolgical invariants are given in \cref{tab:summaryclassif3MA} (codimension 3) and an illustration is provided in \cref{tab:classifAfocalCodim3} .
    %The details of the calculus are given in Appendix~\ref{sec:AfocalCodim3} and an illustration in Table~\ref{tab:classifAfocalCodim3}.

% \textbf{To continue}\\
%     Key arguments:
%     \begin{itemize}
%         \item $2A_1v_2-B_1u_2=u_2P_2+T_1A_2$ with  $P_2=-d_1G(G^2-1)$, $T_1=-2G^2$ and $\Delta =P_2^2+T_2A_2$ with  $T_2=-4G^2d_1(G - 1)^2$.
%         \item $2A_1v_3-B_1u_3=u_3P_3+\wt{T_1}A_3$ with $P_3=-d_1G(G-1)(3G-1)$, $\wt{T_1}=-2G^3$ and $\Delta =P_3^2+T_2A_3$ with  $T_2$ as above.
%         \item $\wt{E_\infty}=\emptyset$
%     \end{itemize}
  
  %\subsection{\label{sec:figures}Figures}

  \section{Conclusion and perspectives}
  
  In this paper, we describe the connected components of the admissible sets associated with three-mirror focal and afocal telescopes that satisfy a set of first-order equations commonly used in on-axis optical design explorations.  %given respectively by \eqref{eq:C1focale}-\eqref{eq:C2Petzval}-\eqref{eq:C3TC} and  \eqref{eq:C1grandissement}-\eqref{eq:C2PetzvalAfoc}-\eqref{eq:C3PupAfoc}. 
  More precisely, we provide a semi-algebraic representation of their connected components and introduce an on-axis nomenclature given in Definition~\ref{def:nomenclature} which serves as a topological invariant for systems with $N\geq 1$ mirrors over the studied admissible set. 
  This latter  is  exact for  all cases summarized in Table~\ref{tab:summaryclassif3MA}.% except for the  focal case of codimension 3 whose two pairs of connected components share the same name ($\texttt{PP010}$ and $\texttt{PP011}$).
    ~As far as our knowledge, it is the first time that such mathematical  aspects of optical solution set are studied.
    Thanks to this mathematical framework, optical designers can rely on the nomenclature defined in Definition~\ref{def:nomenclature} to ensure that all topologically similar optical configurations at the first-order level have been examined, with none overlooked. 
  Furthermore, the semi-algebraic representation allows for faster and more precise sampling of the parameters' space.
  In future work, we aim to examine the case 
  $N=4$, focusing  on the geometry of the solution set for four-mirror focal telescopes.

  \begin{table}[H]    
    \begin{center}
        \begin{tabular}{|c|@{}c@{}|c|}
                      \hline
                      \diagbox[width=3cm,height=0.8cm,dir=SE]{\textbf{codim}}{\textbf{case}}
                       & focal&afocal\\
            \hline
            2& \begin{tabular}{>{\centering\arraybackslash}m{3.2cm}|>{\centering\arraybackslash}m{3.2cm}}
               $f=-1$& $f=1$\\
              \hline
              \begin{tabular}{c}\footnotesize\texttt{NP101,PN101,NP110, }\\ \footnotesize\texttt{PN011}\\\footnotesize$|\pi_0(E)|=4$\end{tabular}
                &\begin{tabular}{c}\footnotesize\texttt{PP010,PP110,PP001,}\\\footnotesize\texttt{PP011,PP100,PP101}\\ \footnotesize$|\pi_0(E)|=6$\end{tabular}\\
            \end{tabular}&
            \begin{tabular}{c}
              \footnotesize\texttt{NP110,NP101,PN101,PN011}\\
              \footnotesize$|\pi_0(E)|=4$
            \end{tabular}\\
            \hline
            3&\begin{tabular}{>{\centering\arraybackslash}m{3.2cm}|>{\centering\arraybackslash}m{3.2cm}}
              $f=-1$&$f=1$\\
              \hline
              \footnotesize \texttt{PN011}&\footnotesize\texttt{PP010,PP110,PP001}\\
              \footnotesize $|\pi_0(E)|=1$&\footnotesize\texttt{PP011,PP101}\\
              &\footnotesize$|\pi_0(E)|=5$\\
            \end{tabular}&
            //
            \\
             \hline           
        \end{tabular}
    \end{center}
    \caption{\label{tab:summaryclassif3MA}Summary of classified topological invariants and number of connected components.%In red, the topological invariants which are shared by two connected components.
     }
\end{table}  
  \section{Aknowledgments}
  The author wants to express its gratitude to N. Tetaz and F. Keller for introducing the optical nomenclature given in Definition~\ref{def:nomenclature},
  in order to give a name to a set of telescopes which are optically similar. 
This
has allowed the author to set the mathematical problem for studying the connected components of the admissible telescopes set as defined in Definition~\ref{def:admissbleSolution}.
Finally, the author wants to thank B. Aymard for the very helpful discussions about transfert matrix formalism. 
  \section{Appendix}
  \label{sec:appendix}
  \begin{appendices}

\section{
Proof of Theorem~\ref{theo:xi1epsContinuousThroughA1epsB1}}
\label{sec:appendixProofContinuity}
\begin{proof}
    In order to show the theorem, we will show that 
    \begin{align*}
        \Gamma(\xi^{(\epsilon)}_{|^c{\W_1^{(\epsilon)}}})\cap E &= E_1\cap\{\epsilon B_1>0\}\\
        \Gamma(\xi^{(\epsilon)}_{|^c{\W_k^{(\epsilon)}}})\cap E &= E_k\cap\{\epsilon p_k<0\}\quad\forall k\in J_n,
        \end{align*}
        where 
\begin{align*}
    ^c{\W_1^{(\epsilon)}}&:=F_\epsilon\cap\mathcal{W}_{A_1}=\mathcal{W}_{A_1}\cap \{\epsilon B_1>0\},\\
^c{\W_k^{(\epsilon)}}&:=F_\epsilon\cap\mathcal{W}_{A_k}=\mathcal{W}_{A_k}\cap \{\epsilon p_k<0\}\quad \forall k\in J_n,\\
E_k &= \{(\mathbf{y},\mathbf{x})\in E,\ \mathbf{y}\in \mathcal{W}_{A_k}\}\quad \forall k\in \{1,\dots,n\}.
\end{align*} 
    We recall that $(\mathcal{H})$ is the set of assumptions described in \cref{hyp:AkC0}.
    We will show that $E^{(\epsilon)} =\{(\mathbf{y},\mathbf{x})\in E,\ \ \epsilon(2A_1x_1+B_1)\geq 0\}= E\cap \Gamma(\xi^{(\epsilon)})=\{(\mathbf{y},\mathbf{x})\in E,\ \mathbf{y}\in F_\epsilon,\ \epsilon(2A_1x_1+B_1)\geq 0\}$ for $\ \epsilon\in \{-1,1\}$.
It is clear that $\xi^{(\epsilon)}$ is continuous on $F_o\backslash\mathcal{W}_\infty$ and 
\begin{equation}
\label{eq:equalityGraphOutsideWinfty}
    E\cap \cup_{\epsilon\in \{-1,1\}}\Gamma(\xi^{(\epsilon)}_{|_{F_o\backslash\mathcal{W}_\infty}})=E\backslash(\cup_{k\in \{1,...,n\}}E_k). 
\end{equation}\
Let $(\mathbf{y_m})_m\in F_o\backslash\mathcal{W}_\infty$ be a sequence such that $\mathbf{y_m}\longrightarrow \mathcal{W}_{A_1}$,
then by setting $\delta_m=A_1(\mathbf{y_m})\longrightarrow 0$, and $\ov{B_1} = \lim_{n\to\infty}B_1(\mathbf{y}_m)$, $\ov{C_1} = \lim_{n\to\infty}C_1(\mathbf{y}_m)$,  $x_1^{(\epsilon)}(\mathbf{y}_m)$ writes as 
    \begin{align*}
        x_1^{(\epsilon)}(\mathbf{y}_m)=\frac{-\ov{B_1}+\epsilon|\ov{B_1}|-\frac{\ov{C_1}}{\ov{B_1}}\delta_m+o(\delta_m/|\ov{B_1}|)}{\delta_m}
    \end{align*}
    which has a limit value equal to $-\frac{\ov{C_1}}{\ov{B_1}}$ iff the condition $(\ov{B_1}=\epsilon |\ov{B_1}|)\land (\ov{B_1}\neq 0)$ is met that is $\epsilon \ov{B_1}>0$ and eventually $\ov{\mathbf{y}}=\lim_{n\to\infty}\mathbf{y}_m\in {^c\mathcal{W}_1^{(\epsilon)}}$.
We deduce that  $\xi^{(\epsilon)}$ is extendable on $^c\mathcal{W}_1^{(\epsilon)}$ by $\xi^{(\epsilon)}(\ov{\mathbf{y}})=(\frac{\ov{C_1}}{\ov{B_1}},-(\frac{\ov{u_kC_1}-\ov{v_kB_1}}{\ov{A_kB_1}})_{k\in J_n})$. On another side 
     \begin{align}
     (\mathbf{y},\mathbf{x})\in E_{1}&\Longleftrightarrow\left\lbrace\begin{aligned}
        \mathcal{G}(\mathbf{y},\mathbf{x})=\textrm{True}\\
     A_1&=0\\
        B_1x_1+C_1&=0\\
        A_kx_k+u_kx_1+v_k&=0\\\end{aligned}\right.\\
        &\overset{\mathcal{H}-((b)+(d))}{\Longleftrightarrow}\left\lbrace\begin{aligned}
            \mathcal{G}(\mathbf{y},\mathbf{x})&=\textrm{True}\\
        A_1&=0\\
        x_1&=-\frac{C_1}{B_1}\\
        x_k&=\frac{u_kC_1-v_kB_1}{B_1A_k}\quad \text{ for } k\in J_n\\
        \end{aligned}\right.
    \end{align}
    We deduce   that  $E_1\cap\{\epsilon B_1>0\}= E\cap \Gamma(\xi^{(\epsilon)}_{|{^c\mathcal{W}_1^{(\epsilon)}}})$.
    By $\mathcal{H}-(d)$ we conclude that
     \begin{equation}
        E_1=E_1\cap \{B_1\neq 0\}=E\cap \cup_{\epsilon\in \{-1,1\}}\Gamma(\xi^{(\epsilon)}_{|\ov{W_1^{(\epsilon)}}}).
        \label{eq:equalityE1}
    \end{equation}\\
Similarly, let $k\in J_n$ and $(\mathbf{y_m})_m\in F_o\backslash\mathcal{W}_\infty$ be such that $\mathbf{y_m}\longrightarrow \mathcal{W}_{A_k}$.
 By $(\mathcal{H})$-(b) $\ov{A_1}=\lim_{n\to\infty}A_1(\mathbf{y}_m)$ does not cancel and $x_1^{(\epsilon)}$ is continuous on a neighbourhood of $\mathcal{W}_{A_k}$. 
 Similarly, $(\mathcal{H})$-(e) enables to say that  $p_k$ does not cancel on a neighbourhood of $\mathcal{W}_{A_k}$.
 By keeping same notations as previously, we get by \cref{le:compatibility} that
\begin{align*}
    B_k^{(\epsilon)}(\mathbf{y}_m)&=\left(\frac{2A_1v_k-u_kB_1+u_k\epsilon\sqrt{\Delta}}{2A_1}\right)(\mathbf{y_m})
    =\left(\frac{u_k p_k+u_k\epsilon\sqrt{p_k^2+q_kA_k}}{2A_1}\right)(\mathbf{y_m})\\
    &=\frac{\ov{u_k}(\ov{p_k}+\epsilon|\ov{p_k}|)+\epsilon\frac{\ov{u_k q_k}}{2|\ov{p_k}|}\delta_m+o(\delta_m/\ov{p_k})}{2\ov{A_1}}\\
    % &=\frac{u_2(r_1+\epsilon|r_1|)+O(\delta/r_1)}{2A_1}
\end{align*}
Hence 
\begin{align*}
    x_k^{(\epsilon)}(\mathbf{y}_m)&=-\left(\frac{B_k^{(\epsilon)}}{A_k}\right)(\mathbf{y_m})= 
    -\frac{\ov{u_k}(\ov{p_k}+\epsilon|\ov{p_k}|)+\epsilon\frac{\ov{u_k q_k}}{2|\ov{p_k}|}\delta_m+o(\delta_m/\ov{p_k})}{2\ov{A_1}\delta_m}\\
\end{align*}
which  has a finite limit value when $\delta_m \longrightarrow 0$ equal to $\frac{\ov{u_k}\ov{q_k}}{4\ov{p_kA_1}}$ iff the condition $(\ov{p_k}=-\epsilon |\ov{p_k}|)\land (\ov{p_k}\neq 0)$ is met that is $\epsilon \ov{p_k}<0$ and eventually $\ov{\mathbf{y}}=\lim_{n\to\infty}\mathbf{y}_m\in {^c\mathcal{W}_k^{(\epsilon)}}$.
Let us compute the limit value of $\xi^{(\epsilon)}(\mathbf{y}_m)$  when $\mathbf{y_m}\longrightarrow \ov{\mathbf{y}}\in{^c\mathcal{W}_k^{(\epsilon)}} $  in such case:
\begin{itemize}
    \item[]\textbf{(i) Case $n=2$.} By \cref{hyp:codim}, $\ov{x_2^{(\epsilon)}}=\frac{\ov{u_2q_2}}{\ov{4p_2A_1}}=-\frac{\ov{A_1C_0}}{\ov{B_1}}$. 
    By $\mathcal{H}-((a)+(c))$ and \cref{hyp:codim} we deduce $\ov{C_1}=0$, and thanks to $\mathcal{H}-(b)$ we get that  $\ov{A_1}\neq 0$, hence $\ov{x_1^{(\epsilon)}}=-\ov{B_1}/\ov{A_1}$  and eventually $\ov{\xi^{(\epsilon)}}=(-\frac{\ov{B_1}}{\ov{A_1}},-\frac{\ov{A_1C_0}}{\ov{B_1}})$. On another side, 
    \begin{align*}
     (\mathbf{y},\mathbf{x})\in E_{2}&\overset{\mathcal{H}-(a)}{\Longleftrightarrow}\left\lbrace\begin{aligned}
        \mathcal{G}(\mathbf{y},\mathbf{x})&=\textrm{True}\\
     A_2&=0\\
        A_1x_1+B_1&=0\\
        x_2x_1&=C_0\\\end{aligned}\right.
        &\overset{\mathcal{H}-((b)+(e))}{\Longleftrightarrow}\left\lbrace\begin{aligned}
            \mathcal{G}(\mathbf{y},\mathbf{x})&=\textrm{True}\\
        A_2&=0\\
        x_1&=-\frac{B_1}{A_1}\\
        x_2=-\frac{A_1C_0}{B_1}\\
        \end{aligned}\right.
    \end{align*}
   We deduce   that  $E_2\cap\{\epsilon p_2<0\}= E\cap \Gamma(\xi^{(\epsilon)}_{|^cW_2^{(\epsilon)}})$.
   \item []\textbf{(ii) Case $n=3$.} Since $\ov{\n A_1p_2}\neq 0$ by $(\mathcal{H})-((b)+(c)+(e)$ we get that $\overline{x_2^{(\epsilon)}}=\frac{\ov{u_2q_2}}{4\ov{A_1p_2}}=\frac{4\ov{u_2A_3C_0A_1}}{4\ov{\n A_1p_2}}=\frac{\ov{u_2A_3C_0}}{\ov{\n p_2}}$.
    Eventually as $\ov{u_2}\neq 0$ by $\mathcal{H}-(f)$, we obtain that $\ov{x_1^{(\epsilon)}}=\frac{-\ov{B_1}+\epsilon|\ov{p_2}|}{2\ov{A_1}}=-\frac{\ov{u_2}(\ov{B_1}+\ov{p_2})}{2\ov{u_2A_1}}\overset{\textrm{\cref{le:compatibility}}}{=}-\frac{\ov{v_2}}{\ov{u_2}}$ and since $\ov{A_3}\neq 0$ by $\mathcal{H}-(b)$ we get $\ov{x_3^{(\epsilon)}}=-\frac{-\ov{u_3}\frac{\ov{v_2}}{\ov{u_2}}+\ov{v_3}}{\ov{A_3}}=\frac{\ov{p_2\n}}{\ov{u_2A_3}}$.
     Hence $\ov{\xi^{(\epsilon)}}=(-\frac{\ov{v_2}}{\ov{u_2}},\frac{\ov{u_2A_3C_0}}{\ov{\n p_2}},\frac{\ov{p_2\n}}{\ov{u_2A_3}})$. On another side, 
   \begin{align*}
     (\mathbf{y},\mathbf{x})\in E_{2}&\Longleftrightarrow\left\lbrace\begin{aligned}
        \mathcal{G}(\mathbf{y},\mathbf{x})&=\textrm{True}\\
        A_2&=0\\
     A_1x_1^2+B_1x_1+C_1&=0\\
        u_2x_1+v_2&=0\\
        A_3x_3+B_3&=0\\
        x_2x_1&=C_0\\\end{aligned}\right.
        &\overset{\textrm{\cref{le:compatibility}}\atop+\mathcal{H}-((b)+(c)+(e))}{\Longleftrightarrow}\left\lbrace\begin{aligned}
            \mathcal{G}(\mathbf{y},\mathbf{x})&=\textrm{True}\\
        A_2&=0\\
        x_1&=\frac{-B_1+\epsilon\sqrt{p_2^2}}{2A_1}=-\frac{v_2}{u_2}\\
        x_3&=-\frac{u_3x_1+v_3}{A_3}=\frac{p_2\n}{u_2A_3}\\
        x_2&=\frac{C_0}{x_3}=\frac{u_2A_3C_0}{p_2\n}\\
        \end{aligned}\right.
    \end{align*}
    We deduce   that  $E_2\cap\{\epsilon p_2<0\}= E\cap \Gamma(\xi^{(\epsilon)}_{|{^c\mathcal{W}_2^{(\epsilon)}}})$.
Similarly we can show that  $E_3\cap\{\epsilon p_3<0\}= E\cap \Gamma(\xi^{(\epsilon)}_{|{^c\mathcal{W}_3^{(\epsilon)}}})$.
\end{itemize}
By $\mathcal{H}-(e)$ we deduce, for both the cases that 
\begin{equation}
    \forall k\in J_n\quad 
    E_k=E_k\cap \{p_k\neq 0\}=E\cap \cup_{\epsilon\in\{-1,1\}}\Gamma(\xi^{(\epsilon)}_{|{^c\mathcal{W}_k^{(\epsilon)}}}).
    \label{eq:equalitEk}
\end{equation}
Finally, merging \eqref{eq:equalityGraphOutsideWinfty}-\eqref{eq:equalityE1}-\eqref{eq:equalitEk} we get that 
$
E = E\cap\left(\cup_{\epsilon\in \{-1,1\}}\Gamma(\xi^{(\epsilon)})\right)
$,
hence $E^{(\epsilon)}\subset E\cap\Gamma(\xi^{(\epsilon)})$ and the other inclusion being evident, we get the equality.
 By introducing $h_\epsilon(\mathbf{y})=(\mathbf{y},\xi^{(\epsilon)}(\mathbf{y}))$ we get that $h_\epsilon:\pi(E^{(\epsilon)})\longrightarrow E^{(\epsilon)}$ is continuous and $h_\epsilon\circ \pi_{|E^{(\epsilon)}} = I_{|E^{(\epsilon)}}$. Since $\pi$ is continuous too, this  shows that $E^{(\epsilon)}\cong \pi(E^{(\epsilon)})$. 
%and as $E\cap\Gamma(\xi^{(\epsilon)})\subset \{(\mathbf{y},\mathbf{x})\in E\ : \ \epsilon( 2Ax_1+B_1)\geq0\}$ which is in fact an equality thanks to the previous equality.
\end{proof} 

\let\clearpage\relax
\nopagebreak
\end{appendices}
  
  % The outline is not required, but we show an example here.
  
 % \section{Main results}
  %\label{sec:main}

  % \bibliographystyle{my_plain}
  % \bibliography{inputs/bibClassif}

%\input{main.bbl}
% \bibliographystyle{plain}
% \bibliography{inputs/bibClassif}
\end{document}